\documentclass[fleqn,usenatbib]{mnras}

\usepackage{newtxtext,newtxmath}
\usepackage[T1]{fontenc}

\DeclareRobustCommand{\VAN}[3]{#2}
\let\VANthebibliography\thebibliography
\def\thebibliography{\DeclareRobustCommand{\VAN}[3]{##3}\VANthebibliography}

\usepackage{graphicx}	% Including figure files
\usepackage{amsmath}	% Advanced maths commands
\usepackage{dblfloatfix}
\usepackage{multicol} 
\title[3D-FireTOC]{Accurate 3D fireball trajectory and orbit calculation using the 3D-FireTOC automatic Python code}

\author[E. Peña-Asensio et al.]{
Eloy Peña-Asensio$^{1,2}$\thanks{E-mail: eloy.pena@uab.cat}, Josep Maria Trigo-Rodríguez$^{1,3}$, Maria Gritsevich$^{4,5,6}$ and Albert Rimola$^{2}$
\\
$^{1}$ Institut de Ciències de l’Espai (ICE, CSIC), Campus UAB, C/ de Can Magrans s/n, 08193 Cerdanyola del Vallès, Catalonia, Spain\\
$^{2}$Departament de Química, Universitat Autònoma de Barcelona, 08193 Bellaterra, Catalonia, Spain\\
$^{3}$Institut  d’Estudis  Espacials  de  Catalunya  (IEEC),  08034  Barcelona,  Catalonia,  Spain\\
$^{4}$Finnish Geospatial Research Institute (FGI), Geodeetinrinne 2, FI-02430 Masala, Finland\\
$^{5}$Department of Physics, University of Helsinki, Gustaf Hällströmin katu 2a, P.O. Box 64, FI-00014 Helsinki, Finland\\
$^{6}$Institute of Physics and Technology, Ural Federal University, Mira str. 19. 620002 Ekaterinburg, Russia
}

\date{Delivered 2020 November 26.}

\pubyear{2020}

\begin{document}
\label{firstpage}
\pagerange{\pageref{firstpage}--\pageref{lastpage}}
\maketitle

\begin{abstract}
The disruption of asteroids and comets produces cm-sized meteoroids that end up impacting the Earth’s atmosphere and producing bright fireballs that might have associated shock waves or, in geometrically-favorable occasions excavate craters that put them into unexpected hazardous scenarios. The astrometric reduction of meteors and fireballs to infer their atmospheric trajectories and heliocentric orbits involves a complex and tedious process that generally requires many manual tasks. To streamline the process, we present a software package called SPMN 3D Fireball Trajectory and Orbit Calculator (\textit{3D-FireTOC}), an automatic \textit{Python} code for detection, trajectory reconstruction of meteors, and heliocentric orbit computation from video recordings. The automatic \textit{3D-FireTOC} package comprises of a user interface and a graphic engine that generates a realistic 3D representation model, which allows users to easily check the geometric consistency of the results and facilitates scientiﬁc content production for dissemination. The software automatically detects meteors from digital systems, completes the astrometric measurements, performs photometry, computes the meteor atmospheric trajectory, calculates the velocity curve, and obtains the radiant and the heliocentric orbit, all in all quantifying the error measurements in each step. The software applies corrections such as light aberration, refraction, zenith attraction, diurnal aberration and atmospheric extinction. It also characterizes the atmospheric flight and consequently determines fireball fates by using the $\alpha - \beta$ criterion that analyses the ability of a fireball to penetrate deep into the atmosphere and produce meteorites. We demonstrate the performance of the software by analyzing two bright fireballs recorded by the Spanish Fireball and Meteorite Network (SPMN).
\end{abstract}


%%%%%%%%%%%%%%%%%%%%%%%%%%%%%%%%%%%%%%%%%%%%%%%%%%

%%%%%%%%%%%%%%%%% BODY OF PAPER %%%%%%%%%%%%%%%%%%

\section{Introduction}

Meteor networks provide valuable scientific information about mm- to m-sized meteoroids crossing the Earth’s orbit thanks to the continuous monitoring of the night sky \citep{Ceplecha1987}. First meteor networks were based on classic photography, but after the first application of CCD and video techniques to meteor observations \citep{trigo2005a, Trigo2006, Madiedo2007} great progress has been made. Nowadays, just a few decades after this digital revolution, CCD and video cameras produce enough meteor recordings to provide an accurate depiction of bright fireballs.

Meteor detection provides information about the origin of meteoroids and about the continuous decay of asteroids and comets, their main parent bodies \citep{Murad2002}. By studying the heliocentric orbits of meteoroids, identifying meteorite-dropping events, and developing the skills to reconstruct their strewn fields, one gains a better understanding of the impact hazard associated with large meteoroids \citep{Jenniskens1998, brown2002flux, Trigo2007, Gritsevich2012consequences, Trigo2017, Sansom2019, Moreno2020, Moilanen2021}. The recovery and the analyses of new meteorites and the study of their dynamic association with comets, asteroids or planetary bodies give new clues on the physical processes delivering space rocks to Earth \citep{Whipple1957, Trigo2007, Jenniskens2008, TrigoRodrguez2009, Trigo2015orbit, Dmitriev2015orbit}. Moreover, the characterization of the atmospheric flight and the study of mechanical properties of the meteoroids contribute to impact hazard assessment \citep{Trigo2006, Safoura2019}. The analysis of cm- to m-sized meteoroids ablating in the Earth's atmosphere gives valuable clues on the delivery of volatiles to Earth by using meteor spectroscopy \citep{Trigo2003, Trigo2004, Trigo2019}, but also is relevant to test their ability to penetrate deep into the atmosphere and quantify the consequences of small asteroids coming from similar sources for planetary defense \citep{brown2002entry,Boslough2008, Silber2018physics}. 

Meteor monitoring differs from most other types of astronomical observations since these luminous events cannot be predicted either in time or space \citep{Vinkovic2020}. For this reason, it is important to monitor the sky with full-time and maximum spatial coverage. That is the foremost goal of multiple stations systems, often referred to as a meteor network.  Some detection networks are tuned to very bright meteors, called fireballs when they exceed the brightness of Venus or superbolides when they are brighter than the magnitude $-16$ \citep{Trigo2015orbit}. Over the years, meteor and fireball detection networks have been built in many parts of the world, for instance, the Harvard Meteor Project \citep{Jacchia1956}, the European Fireball Network \citep{Ceplecha1957}, the continental scale Desert Fireball Network (DFN) \citep{Bland2004}, the SPanish Meteor Network (SPMN) \citep{trigo2005a}, the Southern Ontario All-Sky Meteor Network \citep{Weryk2007}, the Finnish Fireball Network (FFN) \citep{Gritsevich2014}, the French Fireball Recovery and InterPlanetary Observation Network and Meteorite Network (FRIPON) \citep{Colas2014}, the Italian network for meteors observations and trajectory studies (PRISMA) \citep{Gardiol2016} and the Global Fireball Observatory \citep{Devillepoix2020}.

Since 1999 the SPMN has been continuously monitoring the sky over the Iberian Peninsula by setting about 34 stations distributed throughout Spain \citep{trigo2005a, Trigo2006}. All the data used in this work were obtained by the SPMN and the data were processed from the database created and operated by the Meteorite, Minor Bodies and Planetary Sciences group at the Institute of Space Sciences (CSIC-IEEC). The network stations consist of two operational systems: 1) All-sky CCD cameras ($180^{\circ}$) with fish-eye lenses and detectors of $4096x4096$ pixels \citep{trigo2005a}, and 2) wide-field video systems ($90^{\circ}$ to $120^{\circ}$) working at 25 frames per second (transformed into 50 frames per second by deinterlacing) \citep{Madiedo2007}. For the first system, the entire sky can be recorded without interruption and reaching stellar magnitude between $+6$ and $+10$, depending on the zenith angle and the night sky background luminosity. In the case of the second instrumentation, the typical configuration uses 3 cameras per station covering $120x90$ square degrees up to a limiting magnitude of $+4$.

\section{Automatic meteor detection and software analytical procedures}

The astrometric reduction of fireballs involved hitherto a complex and tedious processes that generally required several manual tasks. The future of fireball analysis is oriented towards complete automation of the process, as demonstrated by the latest efforts \citep{Colas2020}. To streamline the study of meteors, we developed a software called 3D Fireball Trajectory and Orbit Calculator (\textit{3D-FireTOC}), an automatic \textit{Python} code for detection, trajectory reconstruction of meteors and heliocentric orbit computation from CCD recordings. Thanks to the rendering engine \textit{VTK} integrated into the \textit{MayaVI} package and using the NASA visible Earth catalogue for rendering the surface (http://visibleearth.nasa.gov/), a realistic 3D model generator was implemented. Furthermore, we developed a friendly graphical user interface based on the toolkit \textit{Qt}.

This software was developed in the framework of the analyses of Spanish Meteor and Fireball Network (SPMN) automatic recordings of fireballs from optical stations distributed across the Iberian Peninsula. The main steps of the analyses are set out below: 1) Meteor trace detection, 2) Star identification and photometry, 3) Pixel to real-world transformation, 4) Atmospheric trajectory reconstruction, 5) Parametrisation of the atmospheric flight, and 6) Heliocentric orbit computation.

Finally, as examples of reduction procedures two meteoric events are discussed: the Taurid Fireball: SPMN251019B and the sporadic superbolide: SPMN160819. The first study case involves typical recordings from video monitoring stations, while the second case involves one recording station, an occasional picture of the trail and data extracted from a US government sensor recording. Using these data the software was able to reconstruct the meteoroid trajectories, computing their masses, luminosities and, from the computation of their radiant and initial velocities, obtaining the heliocentric orbits they had prior to the impact.

\subsection{Fireball trace detection}
\label{sec:maths} 

\begin{figure*}
	\includegraphics[width=2\columnwidth]{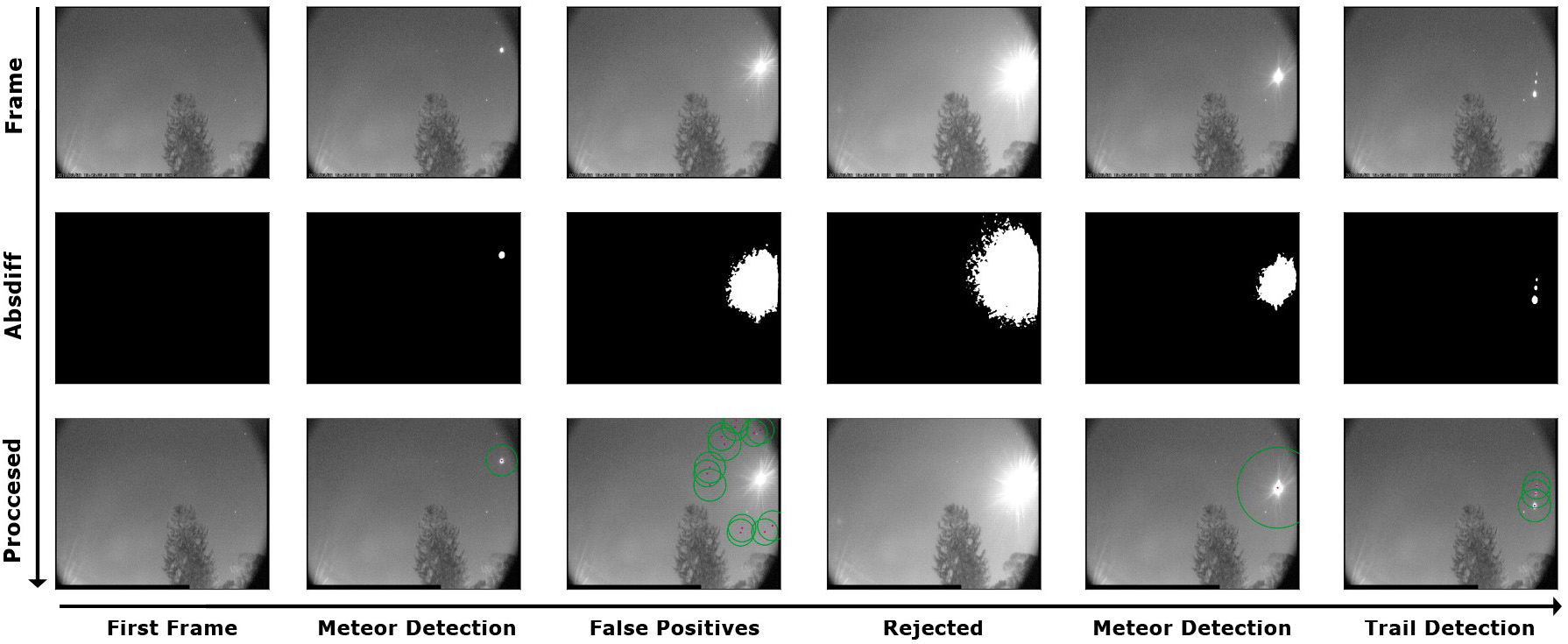}
    \caption{Frames from SPMN300319B fireball, an intermediate step in the processing and the result. Event recorded from Observatorio Astronómico Ramón Maria Aller (OARMA).}
    \label{fig:figure1}
\end{figure*}

A key step in developing an automatic astrometry is ensuring the software's ability to detect the meteors appearing into the field of view of the video detection systems. For this purpose, we used the open-source \textit{CV2 OpenCV} library \citep{opencv_library}. Computer vision techniques are applied to obtain the corresponding pixel coordinates with the meteor moving from frame to frame. Only frames below a fixed mean pixel value will be processed in the most common cases, but saturated frames can appear during the recording of the brightest bolide flares. These sudden increases in the meteor magnitude often saturate the images, so they cannot be properly treated. The first step in processing a video frame uses the typical method of Gaussian blur smoothing of the grayscale image to reduce noise \citep{wells1986efficient}. This enhances image structure and reduces details by convolutions. Each frame is compared to a reference frame (the frame prior to the beginning of the event), so we can extract the pixel value difference above a threshold using the Absdiff function from the CV2 package. To improve the detection, some morphological operations are applied such as erosion (removing isolated pixels) and dilation (expanding the pixel size). The next step is to contour the differenced pixels, as done by \citet{Suzuki1985}, to outline the halo of bright pixels. These contours in subsequent frames can be used to determine if the feature has moved. In each frame, the contour should be the meteoroid or its trail. If the feature is determined to be the meteoroid, the centroid of the contour can be used as the location of the meteoroid (in pixel coordinates). Figure~\ref{fig:figure1} shows a selection of frames from the processed event SPMN300319B, where fireball detections are shown in chronological order from left to right, and their subsequent processing steps top to bottom (see also Table~\ref{table:events} for observer information). The first shown frame precedes the appearance of the meteor. The following frames demonstrate: detection of the meteor, a false positive due to glare, a rejected frame when brightness of the meteor has saturated the image, another detection of the meteor and the detection of the meteor trail.

\begin{figure*}
	\includegraphics[width=2\columnwidth]{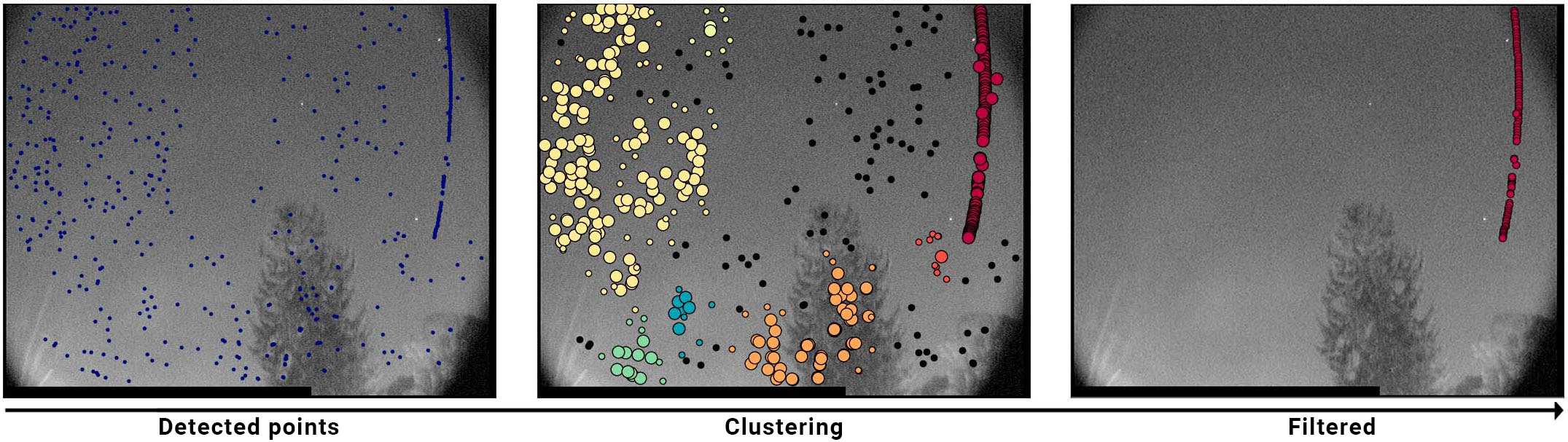}
    \caption{Clustering algorithm and statistical calculations for discarding false positives and automatically selecting the points corresponding to the meteor trail. From left to right: All detected points, clusters found and noise, and selected fireball cluster track. Applied to the SPMN300319B event recorded from OARMA.}
    \label{fig:figure2}
\end{figure*}

Due to the changing nature of fireball recordings, three different methods to avoid false positives are implemented: 1) Discriminating by contour area size excluding excessively small and large contours, 2) Using the first detected points, a Kalman filter predicts an expected area for the next bolide position restricting the contours search \citep{welch1997introduction, Sansom2015} (function implemented in \textit{CV2}). However, if the first detections are not correct, the filter produces wrong predictions, and 3) In parallel, all detected points are saved (including those discarded with the Kalman filter). If the above method does not give a result consistent with a nearly continuous straight trajectory, after the detection process the clustering algorithm DBSCAN is applied to rule out incorrect points \citep{Ester1996}. The cluster associated with the meteor path will present a very low point dispersion, unlike obstacles or glare. See Figure~\ref{fig:figure2}.

It is worth noting that SPMN control computers are synchronised either using GPS controlled systems or using a known software called \textit{NetTime} that guarantees a minimum time accuracy of $0.1\, s$, often slightly better $0.01\, s$. The velocity is determined from the typical $1/25\, s$ frame rate video frequency that could be improved to $1/50\, s$ when deinterlacing the imagery.

\subsection{Star identification and photometry} \label{Photometry}

\begin{figure*}
	\includegraphics[width=2\columnwidth]{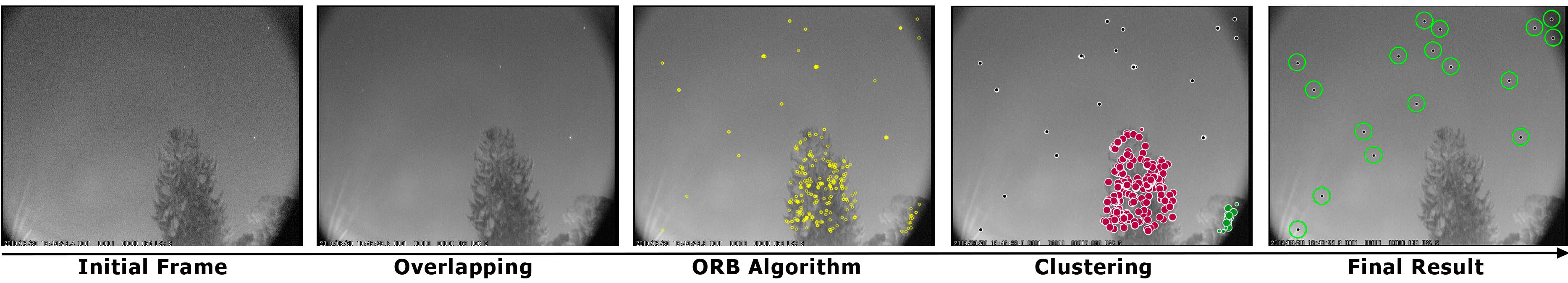}
    \caption{Sequence of the process of obtaining the coordinates of the stars in the image. Depicted temporarily from left to right. It shows the first frame of the video, the overlapping of all valid frames without detection, the application of the ORB algorithm after a logarithmic correction, the classification with the clustering algorithm, and the result. Applied to the SPMN300319B event recorded from OARMA.}
    \label{fig:figure3}
\end{figure*}

To convert pixel coordinates into equatorial coordinates, it is first necessary to identify reference stars with known declination and right ascension. To highlight the stars and reduce noise and spontaneous fireball flashes, all frames without detection are overlapped and a logarithmic correction is applied to improve the process. For the automated identification of the star coordinate on the image, the Oriented FAST and Rotated BRIEF (ORB) descriptor is used \citep{Rublee2011}. Once again, DBSCAN clustering algorithm is used: since stars appear in the sky far from each other in a random distribution, the data labelled as noise by the algorithm will be the one of interest. Figure~\ref{fig:figure3} shows some of the most relevant steps of this process applied to the fireball event SPMN300319B.

It is necessary to apply different corrections to the reference stars that will be used as a comparison to properly estimate the fireball magnitude. The star magnitude found in the catalogue is not the same magnitude as observed from Earth. This is due to the different physical phenomena produced by the atmosphere and the Earth's motion, which must be corrected to obtain proper results. 

The more air mass a star's light passes through to the observer, the more its brightness is reduced due to absorption and scattering processes. We correct this atmospheric extinction using the table made by \citet{Green1992}, based on theoretical values for different atmospheric conditions taking into account the observer altitude and the zenith angle of the star.

The atmosphere of the Earth exhibits a non-uniform optical distribution as a function of altitude. This means that starlights are refracted as their velocity changes from layers with different densities. Because the atmosphere is thin compared to Earth’s radius, it may be treated as a plane parallel to the surface. This simplification allows to easily apply Snell's law and obtain a relation between the true zenith distance $z$ and the apparent zenith distance $\zeta$:
\begin{equation} \label{Snell}
\sin{\zeta}\cos{(z-\zeta)}+\cos{\zeta}\sin{(z-\zeta)} = n\sin{\zeta}.
\end{equation}
Assuming $z-\zeta$ is negligible and dividing by $\cos \zeta$:
\begin{equation} \label{appZenit}
z-\zeta = (n-1)\tan\zeta.
\end{equation}

Although the refractive index at sea level may change with pressure and temperature, we assume that on average $n \approx 1.0003 $. The apparent position of the star can be expressed then as:
\begin{equation} \label{RADErefr}
\begin{split}
\alpha' & = LST - \arctan{\left( \frac{\sin{A}\tan{\zeta}}{\cos{\phi}-\sin{\phi}\cos{A}\tan{\zeta}}  \right)}, \\
\delta' & =\arcsin{ \left( \sin{\phi}\cos{\zeta}+\cos{\phi}\sin{\zeta}\cos{A} \right)},\\
\end{split}
\end{equation}
where $A$ is true azimuth, $\phi$ is the observer's latitude, $LST$ is the local sidereal time, $\alpha'$ is the apparent right ascension, $\delta'$ is the apparent declination \citep{Tatum2019}.

The aberration of light is a phenomenon that occurs due to the vector difference between the velocity of the Earth and the starlight's velocity. This effect displaces the star towards the Earth's apex and may be corrected by using Lorentz transformations:
\begin{equation} \label{Lorentz}
\begin{split}
\cos{\chi'} &= \frac{\cos{\chi}+\frac{v}{c}}{1+\left(\frac{v}{c}\right)\cos{\chi}},  \\
\sin{\chi'} &= \frac{\sin{\chi}}{\gamma \left(1+\left(\frac{v}{c}\right)\cos{\chi} \right)},
\end{split}
\end{equation}
where $\gamma$ is the Lorentz factor $1/\sqrt{1-(v/c)^2}$, $\chi$ is the true apical distance, $\chi'$ is the apparent apical distance, $v$ is the Earth's speed and $c$ is the speed of light.

Assuming $v/c<<1$ and applying trigonometric operations it follows that:
\begin{equation} \label{appRAappDEabe}
\begin{split}
\alpha' &= \frac{-\left(\frac{v}{c}\right)\sin{\psi}\csc{\chi}}{\sin{\chi}\sin{\omega}}+\alpha, \\
\delta' &= \frac{\delta\chi}{\cos{\delta}}\left(-\cos{\psi}\sin{\chi}+\sin{\psi}\cos{\omega}\cos{\chi} \right)+\delta,
\end{split}
\end{equation}
where $\alpha$ is the true right ascension, $\delta$ is the true declination and $\omega$ is the angle between the Earth's apex, the star, and the north polar.

Each star moves in the field of view with a relative speed depending on its declination. This causes the stars near the poles to have smaller angular velocities and, hence, to activate the pixels longer. Consequently, as suggested by \citet{Rendtel1993} it is appropriate to correct the magnitude of the stars to a reference declination as follows:
\begin{equation} \label{MagnitudeVel}
m(\delta_s)=m(\delta_{0^\circ})-2.5p\log\frac{1}{\cos{\delta_s}},
\end{equation}
where $p$ is the Schwarzschild exponent (typically between $0.7$ and $0.8$), $\delta_s$ is the star declination, $\delta_{0^\circ}$ is the reference declination and $m$ is the apparent magnitude.

Once the magnitudes of the reference stars are corrected, we perform an aperture photometry by counting the pixel intensity of both the stars and the fireball. In this way, a logarithmic regression can be made to obtain the magnitude of the fireball. To standardize the luminosity, we corrected its magnitude as if it had been observed at the zenith and calculated its absolute magnitude, i.e., its magnitude at 100 km distance:
\begin{equation} \label{AbsMag}
M = m-5\log\frac{h}{100},
\end{equation}
where $h$ is the height and $M$ is the absolute magnitude.

\subsection{Pixel to real-world transformation}

A key process concerns the transformation from the digital chip system coordinates into the equatorial coordinates that we will carry out by comparing reference stars in the detecting field of view (FOV). Once the apparent positions of the reference stars are known, together with their pixel positions, a transformation matrix is computed to convert the plate coordinates $(x, y)$ into equatorial ones $(\alpha, \delta)$. This yields the apparent trajectory of the fireball from each station. However, since the optical system introduces a lens distortion and possible misalignments, there is a displacement of the stellar images from the plate centre. So, pixels cannot be converted directly to equatorial coordinates. Therefore, the transformation needs intermediate steps. It is necessary to transform measured plate coordinates or pixel coordinates into standard coordinates ($\xi$, $\eta$), that is, stereographic projected coordinates or true coordinates, since:
1) The stereographic projection and the camera sensor are not necessarily aligned, 
2) The photographic objective provides a distorted image of the celestial sphere, as it is the result of the sphere projection on the focal plane, and 
3) The wide-field and fish-eye lenses typically produce pincushion or barrel distortion.

After the first empirical proposal of an absolute astrometric model by \citet{Ceplecha1987}, some refinements and improvements to the parameter estimation were suggested \citep{Borovivcka1992astrometry, Borovicka1995new}. However, these models present a high non-linearity, so they are hardly reversible and the convergence of estimation algorithms is not easily achieved. For this reason, new parametrisation based on polynomial representation was proposed \citep{bannister2013numerical, barghini2019astrometric, jeanne2019calibration}. Following these latest works and since SPMN's stations are equipped with very diverse lenses, we model, as a first approximation, the distortion due to the lens with a quadratic expression as suggested by \citet{Hawkes1993}, which can be expanded to higher orders if the number of reference stars allows it:
\begin{equation} \label{Tatum}
\begin{split}
\xi-x &= ax^2+hxy+by^2+gx+fy+c,\\
\eta-y &= a'x^2+h'xy+b'y^2+g'x+f'y+c',
\end{split}
\end{equation}
where $a$, $h$, $b$, $g$, $f$, $a'$, $h'$, $b'$, $g'$, $f'$ are parameter to fit and the plate constants.
Finally, the transformation of standard coordinates into equatorial coordinates \citep{Steyaert1990} is computed: 
\begin{equation} \label{standard2equa}
\begin{split}
\alpha & = A +\arctan \left(\frac{\xi}{\eta\cdot\sin{D}-\cos{D}} \right),\\
\delta & = \arctan \left(\frac{\eta\cdot\cos{D}+\sin{D}}{\sqrt{\xi^2+(\eta\cdot\sin{D}-\cos{D})^2}} \right).
\end{split}
\end{equation}

being $\alpha$ the right ascension, $\delta$ the declination and $(A, D)$ the unknown optical axis.

Since there is no analytical method to find the position of the optical axis, the simplex algorithm is used to find the solution for the system \ref{Tatum} and \ref{standard2equa} that minimizes the mean squared error \citep{motzkin1952new}.

\subsection{Atmospheric trajectory reconstruction}

\begin{figure}
	\includegraphics[width=\columnwidth]{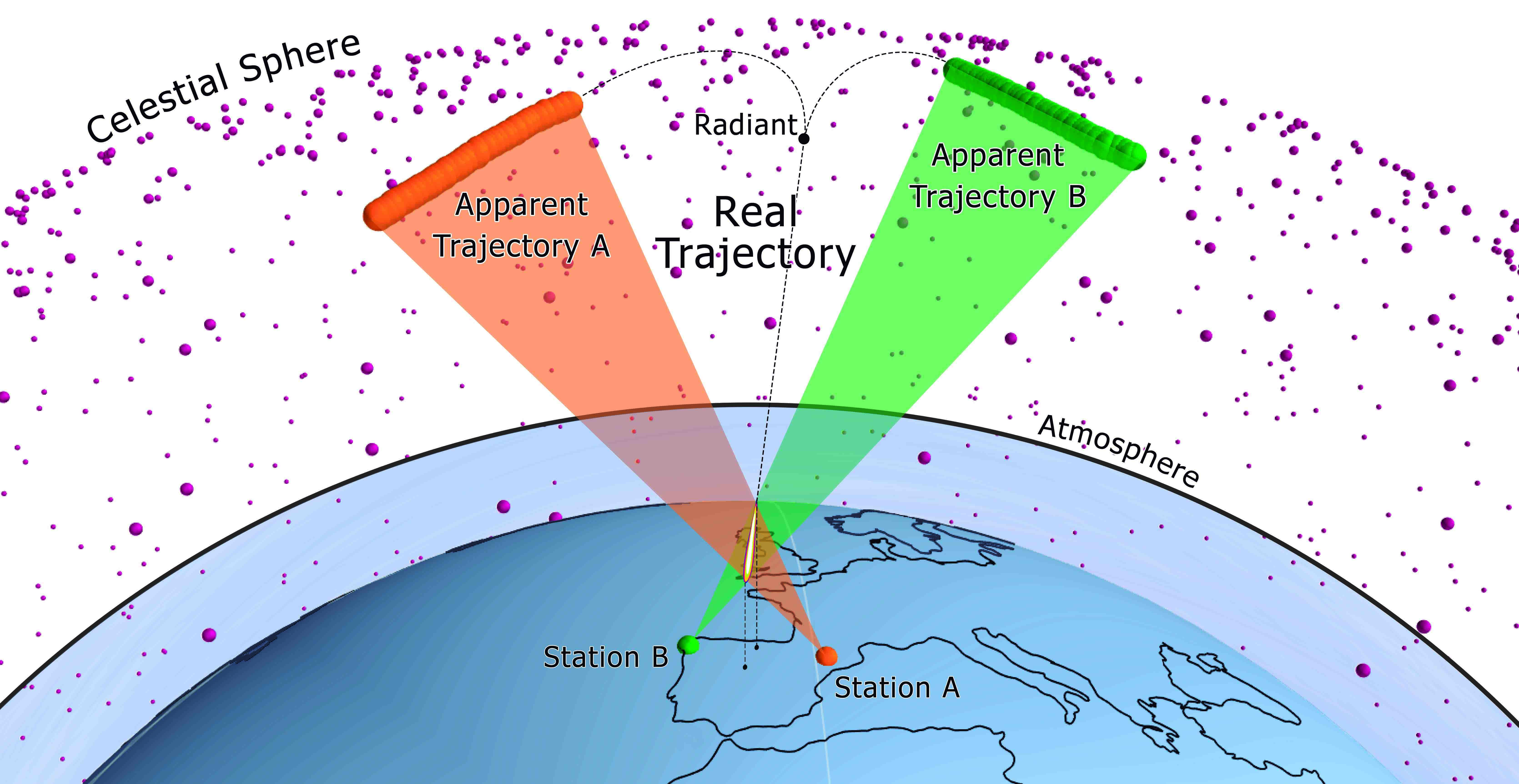}
    \caption{Graphical representation of the real meteor trajectory calculation by intersecting the planes and obtaining the radiant by projecting backwards until the collision with the celestial sphere. The vertical projection is shown as well.}
    \label{fig:figure4}
\end{figure}

Numerous methods have been proposed for meteoroid triangulation \citep{Ceplecha1987, borovicka1990comparison, gural2012new}, some of them very recent \citep{jansen2020dynamic}. We follow the Method of Planes proposed by \citet{Ceplecha1987}, the average plane containing the apparent trajectory and the geographic coordinate from each observation point is obtained and then the stereoscopic intersection of the apparent trajectories calculated. See Figure~\ref{fig:figure4}.

We remark that the radiant is computed, as usual in related bibliography, by doing a backward projection of the atmospheric trajectory from each station until their encounter with the celestial sphere at $R_{\infty}$. Once the first meteor point is known $(X_1, Y_1, Z_1)$ and using any other point contained on the trajectory $(X_2, Y_2, Z_2)$, the radiant can be computed by a line-sphere intersection:
\begin{equation} \label{line2sphere}
\begin{split}
a_R & = (X_2-X_1)^2 + (Y_2-Y_1)^2 + (Z_2-Z_1)^2, \\
b_R & = 2\cdot \big( (X_2-X_1)\cdot X_1 + (Y_2-Y_1)\cdot Y_1 + (Z_2-Z_1)\cdot Z_1 \big), \\
c_R & = X_1^2 + Y_1^2 + Z_1^2 - R_{\infty}^2\\
\end{split}
\end{equation}
being $a_R$, $b_R$ and $c_R$ the parameters of the resulting equation from substituting the equation of the line into the sphere.

This yields to the parametric line:
\begin{equation} \label{line2sphere2}
\begin{split}
t_R & = \frac{-b_R-\sqrt{b_R^2-4 \cdot a_R\cdot c_R}}{2a_R},\\
X_R & = X_1 + t_R\cdot(X_2-X_1),\\
Y_R & = Y_1 + t_R\cdot(Y_2-Y_1),\\
Z_R & = Z_1 + t_R\cdot(Z_2-Z_1),\\
\end{split}
\end{equation}
where the negative root of $t_R$ is chosen since it is the closest point to $(X_1, Y_1, Z_1)$. The cartesian coordinates of the radiant are $(X_R, Y_R, Z_R)$.

The presence of the Earth's gravity not only disturbs the velocity of the meteoroid but also modifies its velocity vector, having consequences in the determination of its radiant in the sky \citep{Dmitriev2015orbit}. The method proposed by \citet{Andreev1990} corrects this shift of the radiant towards the zenith, the so-called zenith attraction. Furthermore, the diurnal aberration has to be taken into account. Since the Earth rotates around its axis, the position of the radiant moves away. The diurnal aberration is caused by the velocity of the observation point on the rotating surface of the Earth. Therefore, it depends not only on the moment at which the observation is made, but also on the latitude and longitude of the observer as the Earth's rotation around its axis moves the position of the radiant as well. It is corrected using the approximation suggested by \citet{Bellot1992}.

Likewise, by performing geometric operations, the height of the meteoroid $h$, the distance to each station, the length travelled and the angle between the fireball trajectory and the local horizon $\gamma$ can be deduced.

\begin{figure}
	\includegraphics[width=\columnwidth]{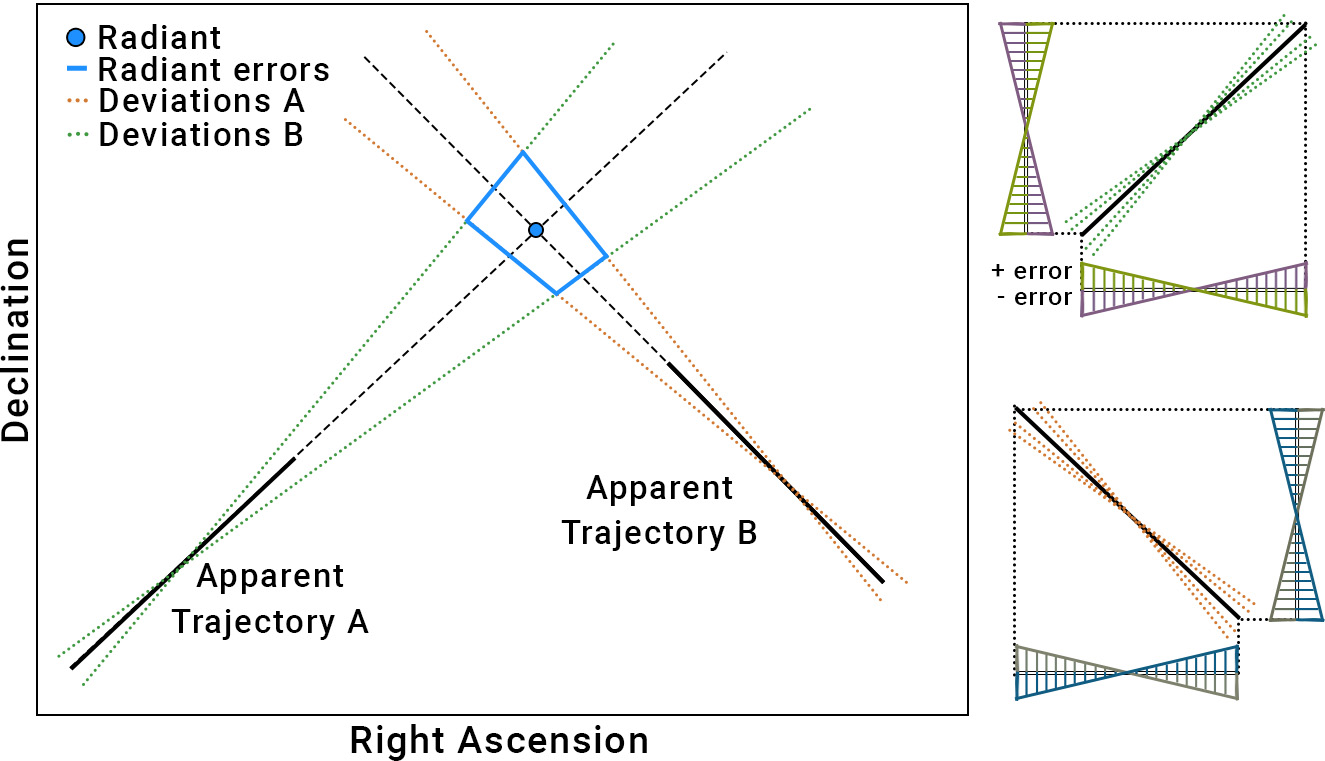}
    \caption{A schematic diagram for radiant error computation. }
    \label{fig:figure5}
\end{figure}

The calculation of errors consists of assuming the worst scenario from the simplex method uncertainties, that is to say, that each point of the apparent trajectory arranges in the way that most deviates from the radiant. This will occur when the points are aligned crosswise along the path, as shown in Figure~\ref{fig:figure5}: on the right of that figure, the four possibilities of deviation assuming the worst case for each right ascension and declination are depicted; on the left, the two largest possible deviations for each apparent trajectory are shown, which delimits the radiant error. The standard deviation assumed in the apparent trajectory comes from how our pixel to real-world transformation matches the reference stars.

In a similar way, using the clone trajectories for the worst-case scenario and following the plane intersection method, we obtain the average observed velocity in the first $10\, \%$ of the luminous trajectory for each of them, as suggested by \citet{Whipple1957}. Velocities are calculated using the closest and/or most reliable observation. Then, we perform a linear regression to estimate the pre-atmospheric velocity at an instant prior to the first detection, specifically, the time interval corresponding to one frame. In this way, we obtain the meteoroid velocity at atmospheric impact and its associated maximum error. Using this velocity, we apply the aforementioned diurnal aberration and zenith attraction corrections propagating the error by deriving the equations involved and taking into account the astrometric errors.

\subsection{Parametrisation of the atmospheric flight}\label{AtmFlight}
One of the most complex parts of the meteor reduction analysis is to develop a mathematical model that properly describes the atmospheric flight, allowing us to extract physical information. Following the classical dynamic third-order time dependent system for characterizing meteor deceleration and assuming that the body does not suffer any kind of fragmentation \citep{Hoppe1937} developed the well-known Single Body Theory (SBT). We adopted this SBT approach which treats the heat exchange and drag coefficients as constants along the luminous flight. 

By introducing convenient dimensionless quantities, the trajectory equations acquire the form of \citep{Gritsevich2009determination}:

\begin{equation} \label{motionflighsimple}
\begin{split}
m\frac{dv}{dy} & = \frac{1}{2}c_d\frac{\rho_0h_0S_e}{M_e}\frac{\rho vs}{\sin{\gamma}},\\
\frac{dm}{dy} & = \frac{1}{2}c_h\frac{\rho_0h_0S_e}{M_e}\frac{V_e^2}{H^*}\frac{\rho v^2s}{\sin{\gamma}},\\
\end{split}
\end{equation}
the scale factor $h_0=7.16\, km$, sublimation heat $H^*$, atmospheric density near the sea level $\rho_0=1.29 \cdot 10^3\, g/cm^3$, dimensionless mass $m=M/M_e$, velocity $v=V/V_e$, air density term $\rho=\rho_a / \rho_0$ and cross-sectional area $s=S/S_e$. The subscript "$e$" indicates the parameters at the entry to the atmosphere.

To find an analytical solution, it is assumed the isothermal atmospheric model $\rho=e^{-y}$ and according to \citet{Levin1956} the body mass and its middle section are connected by introducing the shape change coefficient $s=m^\mu$. The dimensionless parameter $\mu$ is treated as a constant and can be inferred in each case by studying the meteor light curve \citep{Gritsevich2011, Bouquet2014, drolshagen2020luminous}. The first integrals for the system (Eq. \ref{motionflighsimple}) was proposed by \citet{Stulov1995} with the initial conditions $m=1$, $v=1$, $y=\infty$:

\begin{equation} \label{motionflighsol}
\begin{split}
& m = exp \left (  -\frac{\beta}{1-\mu}(1-v^2)  \right ),\\
& y = \ln{\alpha} + \beta- \ln{\frac{\Delta}{2}},\\
\end{split}
\end{equation}
where:
\begin{equation} \label{motionflighsol2}
\begin{split}
& \Delta = \overline{E}i(\beta) - \overline{E}i(\beta v^2), \\
& \overline{E}i(x) = \int_{-\infty}^{x} \frac{e^tdt}{t} dx,\\
\end{split}
\end{equation}
showing that the trajectory depends on two dimensionless parameters:
\begin{equation} \label{alphabeta}
\begin{split}
\alpha & = \frac{1}{2}c_d \frac{\rho_0h_0S_e}{M_e\sin{\gamma}}, \\
\beta & = (1-\mu)\frac{c_hV_e^2}{2c_dH^*}. \\
\end{split}
\end{equation}
where, in this section only, $\alpha$ symbol refers to the ballistic coefficient and $\beta$ to the mass loss parameter.

The parameter $\alpha$ characterizes the aerobraking efficiency since it is proportional to the mass of a trajectory-aligned atmospheric column of cross-section divided by the body mass. The parameter $\beta$ is proportional to the fraction of the kinetic energy supplied to a unit mass of the body as heat divided by the effective destruction enthalpy.

These parameters bring great simplicity to the characterization of the atmospheric flight and can also be used to estimate how likely a fireball produces meteorites \citep{Gritsevich2008pribram, Gritsevich2008estimating, Gritsevich2009classification, Gritsevich2012consequences, Turchak2014meteoroids, Sansom2019, Moreno2020}. In this regard, we implemented latest modification of the method proposed by \citet{Sansom2019} for determining fireball fates using $\alpha - \beta$ criterion. Physically meaningful parametrisation of the luminous flight allows the pre-atmospheric and final mass (corresponding to the terminal height) to be computed:
\begin{equation} \label{Masse}
\begin{split}
M_e &= \left ( \frac{1}{2}\frac{c_d A_e \rho_0 h_0}{\alpha \rho_m^{2/3} \sin{\gamma}}\right )^3, \\
M_f &= M_e \exp{\left ( -\frac{\beta}{1-\mu} \left( 1-\left(\frac{V}{V_e}\right)^2  \right) \right )}, \\
\end{split}
\end{equation}
where $\rho_m$ meteoroid bulk density and $A_e$ is the pre-atmospheric shape factor (usually ranges between 1.21, for an ideal sphere, and 1.8) \citep{Trigo2015orbit, Meier2017, Gritsevich2017, lyytinen2016implications}.

From the initial mass and approximating the shape of the meteoroid to a sphere, the initial size can be estimated. This value can be contrasted with the calculation of the diameter from the radiated energy. Assuming that the kinetic energy value is the registered impact energy $T_E$, the equivalent meteoroid diameter $D$ is computed as:
\begin{equation} \label{Diamater}
D = 2 \sqrt[3]{\frac{3T_E}{2\pi\rho v^2}},
\end{equation}
where $\rho$ is the meteoroid bulk density and $v$ the velocity of the meteor.

It is worth noting that since observed velocities have marked inaccuracies, data must be pre-processed before it can be used to fit the parameters and compute velocities and deceleration. Experience says that an optimal way to approximate these velocities is to adjust the distances with a least-squares to the following equation:

\begin{equation} \label{length}
L = a + bt+ ce^{kt},
\end{equation}
where $L$ is the path length, $t$ is the time and $a$, $b$, $c$ and $k$ are variables to be determined in the curve fitting \citep{Whipple1957, mccrosky1968special}. Once the adjustment is made, by deriving the previous expression velocities and decelerations are obtained in a trivial way.

\subsection{Heliocentric orbit computation}
The last step to know the origin of the meteoroid in the Solar System is to reconstruct its heliocentric orbit. Once the radiant has been obtained and corrected and the atmospheric flight velocity curve has been computed, the orbital elements that define the meteor's orbit can be calculated.
Following the steps of \citet{Ceplecha1987} and \citet{Jenniskens1987}, first the coordinates of the geocentric radiant $(\alpha_G, \delta_G)$ are transformed into ecliptical longitude and latitude $(L_G, B_G)$. Thus, the heliocentric ecliptic system of rectangular coordinates can be defined as:
\begin{equation} \label{eq1}
\begin{split}
X & = r \cdot \cos{L} \cdot \cos{B},  \\
Y & = r \cdot \sin{L} \cdot \cos{B}, \\
Z & = r \cdot \sin{B},
\end{split}
\end{equation}
where $r$ is the distance to the Sun.

Then the ecliptical longitude of the Earth's Apex $L_{AP}$ and the Earth's velocity $V_{AP}$ are extracted from $JPL \,  Horizons$ ephemerides. The heliocentric velocities of the meteoroid can be expressed as:

\begin{equation} \label{eq2}
\begin{split}
v_{H_x} & = - v_G \cdot \cos{L_G} \cdot \cos{B_G} + V_{AP}\cos{L_{AP}},  \\
v_{H_y} & = - v_G \cdot \sin{L_G} \cdot \cos{B_G} + V_{AP}\sin{L_{AP}},\\
v_{H_z} & = - v_G \cdot \sin{B_G}.
\end{split}
\end{equation}

The specific angular momentum $ \bar{h} = (h_x, h_y, h_x)$ and the ascending node vector $\hat{n}$ are needed to determine the parameters, which can be computed as:
\begin{equation} \label{eq3}
\begin{split}
\bar{h} & = \bar{r} \times v,  \\
\bar{n} & = \hat{z} \times \bar{h} = (-h_y, h_x, 0), 
\end{split}
\end{equation}
where $\hat{x}$,$\hat{y}$ and $\hat{z}$ are the unit axes of the heliocentric coordinate system. 

Finally, each of the orbital elements can be obtained:
\begin{equation} \label{eq4}
\begin{split}
& \cos(i) =  \frac{h_z}{h},  \\
&\cos(\Omega) =  \frac{h_y}{\sqrt{h_x^2 + h_y^2}}, \\
& \cos(\omega) =  \frac{-h_y e_x + h_x e_y}{e\sqrt{h_x^2 + h_y^2}}, \\
& \bar{e} = \frac{1}{GM} \left ( \left(v^2-\frac{GM}{r} \right )\bar{r}-(\bar{r}\cdot \bar{v}) \bar{v} \right ), \\
& a = \frac{1}{\frac{2}{r}- \frac{v^2}{GM}}, \\
& \cos{v_0} = \frac{\bar{e}\cdot \bar{r}}{er},
\end{split}
\end{equation}
where $i$ is the inclination, $\Omega$ the longitude of the ascending node, $\omega$ the argument of perihelion, $e$ the eccentricity, $a$ the semimajor axis and $v_0$ the true anomaly \citep{dubiago1961determination}.

\section{Study cases}
The software was successfully applied to study different events as test cases. We tested earlier the computer vision system with the SPMN300319B case as it presented notable complications such as obstacles and frame saturation. We chose two other events to exemplify a complete reduction (see Table~\ref{table:events}). 

\begin{table*}
 \caption{Table with the different events recorded by the SPMN network. The SPMN300318 event is from single station since it was used to illustrate the operation of the code. $^*$The observation point does not belong to the SPMN network.}
 \label{table:events}
 \begin{tabular}{lccccccc}
  \hline
  Name & Stations & Longitude & Latitude & Altitude & Date &  Start Time (UTC) & End Time (UTC)\\
  \hline \hline 
  SPMN300319B & OARMA & 08$^\circ$33'19''W & 42$^\circ$52'33''N & 236 m & 2019/03/30 & 19h46m30.4s & 19h46m34.4s \\
  \hline 
  SPMN251019B & Eivissa & 01$^\circ$25'45''E & 38$^\circ$54'21''N & 45 m & 2019/10/25 & 04h36m48.4s & 04h36m50.4s \\
  & Folgueroles & 02$^\circ$19'33''E & 41$^\circ$56'31''N & 580 m & & 04h36m49.976s & 04h36m50.657s\\
  & Montseny & 02$^\circ$32'01''E & 41$^\circ$43'47''N & 194 m & & 04h36m46.279s & 04h36m48.310s\\
  \hline 
  SPMN160819 & Eivissa & 01$^\circ$25'45''E & 38$^\circ$54'21''N & 45 m & 2019/08/16 & 20h36m01.3s & 20h36m05.6s\\
  & Costa Brava$^*$ & 03$^\circ$04'10''E & 41$^\circ$49'03''N & 2 m & & 20h36m04s & 20h36m04s\\
  & Sardinia$^*$ & 08$^\circ$31'43''E & 39$^\circ$54'37''N & 30 m & & 20h36m01s & 20h36m06s\\

  \hline
 \end{tabular}
\end{table*}

The SPMN251019B fireball is a typical reduction case thanks to the favorable astrometry made based on the recordings from the three stations, which we propose as belonging to the Taurids complex. The other studied case is the superbolide SPMN160819 that demonstrates the ability to combine satellite data and video recordings. These events are listed in Table~\ref{table:events} with their corresponding observation data.

\subsection{Taurid Fireball: SPMN251019B}

The first example is the bolide SPMN251019 that occurred on October 25th, 2019 at 04:36:46 UTC \citep{Eloy2020a}. The event was videotaped by three SPMN monitoring stations: Astronomical observatory at Puig des Molins (Eivissa), Montseny Astronomical Observatory (Barcelona) and Folgueroles (Barcelona). The station coordinates are listed in Table~\ref{table:events}.

One of the complications of this case is that the two closest stations recorded the beginning and the end of the fireball, but not the intermediate part, which was only filmed from the Eivissa station. Despite this, from the astrometric measurements of the video frames and the integration of the data we achieved the trajectory reconstruction. The fireball light was first detected at a height of $79.0 \pm 0.1\,km$ and the end occurred at $58.3 \pm 0.1 \, km$ having a trajectory angle of $\gamma = 28.7\,^{\circ}$, which indicates the very remote possibility of being a meteorite-dropper since its terminal height was too high \citep{Moreno2015new}. Following the photometry procedure described in Section \ref{Photometry}, we obtained a magnitude of $-13.5 \pm 0.5$, as bright as the full moon.

The pre-atmospheric velocity was retrieved from the velocity measured at the earliest part of the fireball trajectory by doing a regression and extrapolating with a backpropagation. It was estimated to be $28.0 \pm 0.2\, km/s$. Assuming a shape change coefficient of $\mu = 2/3$, a shape factor of $A_e = 1.3$, a drag coefficient of $c_d = 1.3$ and a relatively low density of $\rho=1.6 \, g/cm^3$ \citep{HARMON2005}, the initial and final mass were computed using the method detailed in Section \ref{AtmFlight}. Using the D-criterion of \citet{southworth1963statistics} we obtained a value of $D_{SH}=0.35$. This indicates that the orbit of the event SPMN251019B is suggestive of being dynamically associated with the established Southern Taurid shower \citep{Jenniskens2016}.

\begin{figure}
	\includegraphics[width=\columnwidth]{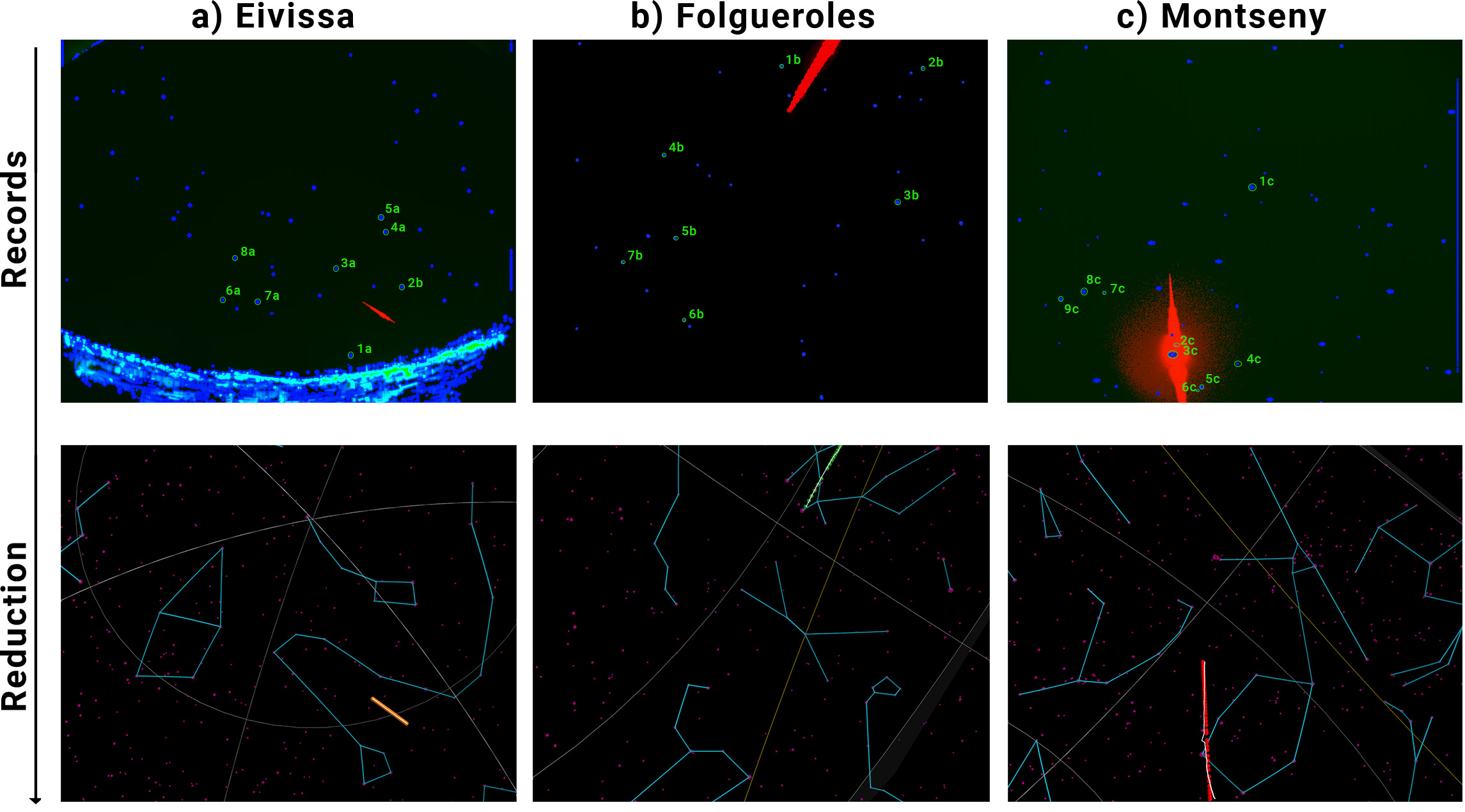}
	\includegraphics[width=\columnwidth]{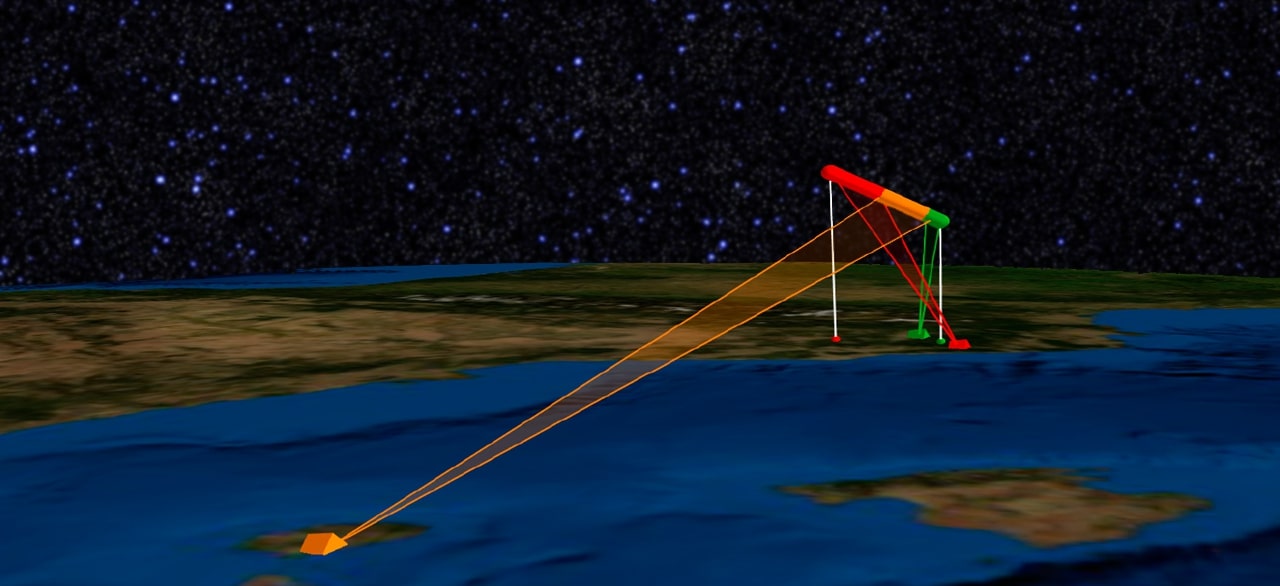}

    \caption{Top: SPMN251019B apparent trajectory recorded and reduced from Eivissa (orange), Folgueroles (red) and Montseny (green). Bottom: SPMN251019B atmospheric trajectory with vertical projection (white).}
    \label{fig:figure6}
\end{figure}

\begin{figure}
	\includegraphics[trim=80 50 80 50 ,clip,width=\columnwidth]{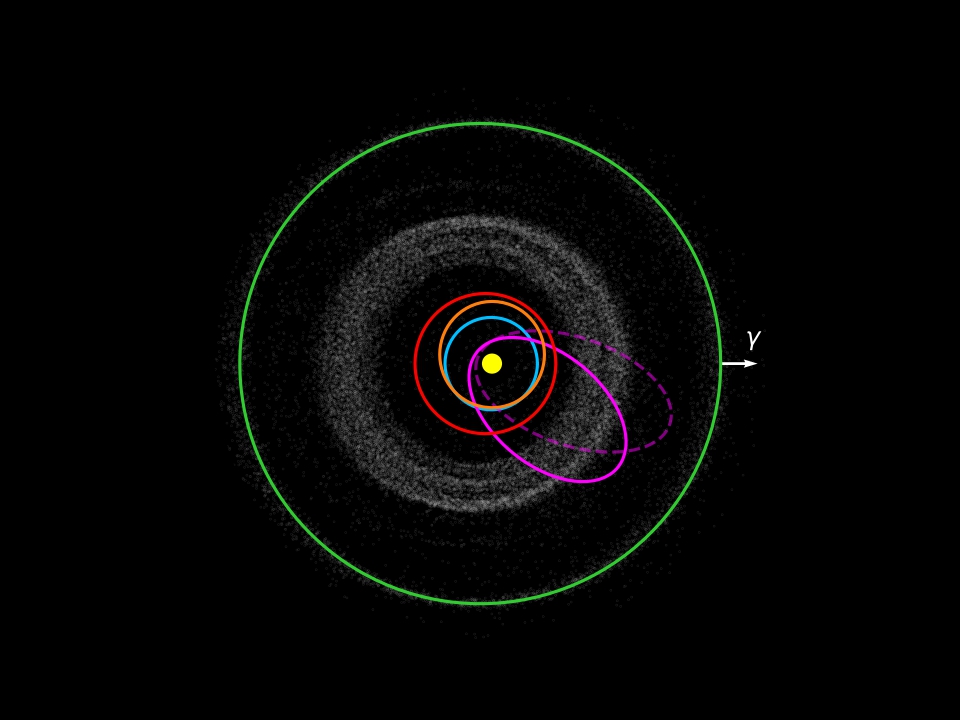}

    \caption{An orbital projection to the ecliptic plane of different solar system bodies is shown, indicating the Vernal equinox to the right. The Sun is shown in yellow, Earth's orbit in blue, Mars's orbit in red, the Main Belt Asteroids in gray, Jupiter in green, the orbit of the SPMN251019B meteoroid in pink, comet 2P/Encke in dashed pink and the orbit of the SPMN160819 meteoroid in orange colour.}
    \label{fig:figure6b}
\end{figure}

The calculated radiant and the velocity are shown in Table~\ref{table:SPMN251019B}, together with the computed orbital parameters and the main fireball parameters. Figure~\ref{fig:figure6} shows the summed frames of the recordings and the graphic representation of the apparent trajectories over the celestial sphere and the atmospheric flight in real scale. In addition, Figure~\ref{fig:figure6b} depicts the orbit of the SPMN160819B progenitor associated to comet 2P/Encke.

Bright fireballs recorded in October are often belonging to one of the Taurid streams (Northern or Southern branches). The Taurids exemplify impact hazard associated with large meteoroids due to the frequency and size of the bodies reaching the Earth’s atmosphere. The entire Taurid complex consists of Near Earth Objects (NEOs), plus several meteoroid streams. The complex itself is considered a potential source of risk related to possible impacts by cosmic objects. In fact, it was proposed that the Tunguska event was produced by an asteroid-size body associated with the Taurid complex \citep{Sekanina1998}. Several studies have demonstrated the dynamic association between the Taurid complex and the disruption of a much larger 2P/Encke progenitor comet (see e.g. \citep{Kresak1978}).

\begin{table}
\scriptsize
\centering
\begin{tabular}{l c c c}
\hline                         
\multicolumn{4}{c}{Radiant Data} \\ 
\hline \hline      
& Observed & Geocentric & Heliocentric \\ 
$\alpha\,$ ($^{\circ}$) & 42.7$\pm$0.2 & 40.5$\pm$0.2 & 346.7$\pm$0.4 \\
$\delta\,$ ($^{\circ}$) & 11.3$\pm$0.1 & 9.5$\pm$0.2 & -4.2$\pm$0.3 \\
V\, (km/s) & 28.0$\pm$0.2 & 26.0$\pm$0.2 & 36.5$\pm$0.2 \\
\hline
\end{tabular}

\scriptsize
\begin{tabular}{l c c c c c}
\hline                         
\multicolumn{6}{c}{Orbital Parameter} \\ 
\hline \hline      
a\, (AU) & e & q & $\omega\,$ ($^{\circ}$) & $\omega\,$ ($^{\circ}$) & i$\,$ ($^{\circ}$) \\
\hline
1.97$\pm$0.07 & 0.792$\pm$0.007 & 0.410$\pm$0.006 & 109.2$\pm$0.9 & 31.199$\pm10^{-4}$ & 6.0$\pm$0.4 \\
\end{tabular}

\scriptsize
\begin{tabular}{p{0.89cm} p{0.78cm} p{0.8cm} p{0.88cm} p{0.88cm} p{0.65cm} p{0.65cm}}
\hline                         
\multicolumn{7}{c}{SPMN251019B} \\
\hline \hline      
Mag & $h_i\, (km)$ & $h_f\, (km)$ & $V_i \, (km/s)$ & $V_f \, (km/s)$ & $M_i \, (g)$ & $M_f \, (g)$ \\
\hline
-13.5$\pm$0.5 & 80.0$\pm$0.1 & 58.3$\pm$0.1 & 28.0$\pm$0.2 & 17.59$\pm$0.2 & 43.1 & $\,\,\,$0.003 \\
\hline
\end{tabular}

\caption{Top: SPMN251019B observed, geocentric and heliocentric radiant and velocities. Middle: SPMN251019B calculated orbital parameters. Bottom: SPMN251019B computed atmospheric trajectory, velocity and mass.}
\label{table:SPMN251019B}
\end{table}

\subsection{Sporadic superbolide: SPMN160819}

On August 16, 2019, a very bright superbolide catalogued as SPMN160819 event occurred (see Table \ref{table:events}). It was an event of considerable importance due to its magnitude that, unfortunately, was only partially recorded from the Eivissa station of the SPMN network \citep{Eloy2020b}. However, thanks to citizen collaboration, we had access to two more records: an image from Costa Brava and a video from Sardinia, which were used in the superbolide analysis.

Since casual records of extremely rare events have limited resolution, we had to use the peak brightness coordinates measured by the Center for Near Earth Object Studies (CNEOS) at NASA to perform a correct reduction of this event.

From the recording from Eivissa, in which the Moon appears at a similar altitude, the superbolide was more luminous than the Moon. It was estimated to exhibit an absolute magnitude of $-16.5 \pm 0.5$. The superbolide from Eivissa was so distant that the first detected light was at a height of $67 \pm 3\, km$ and ended at $23 \pm 2 \, km$. The result for the pre-atmospheric velocity was $15 \pm 1 \, km/s$ and the terminal velocity $11 \pm 1 \, km/s$. The pre-atmospheric velocity is consistent with that recorded by CNEOS ($14.9\, km/s$).

\begin{table}
\centering
\scriptsize
\begin{tabular}{l c c c}
\hline                         
\multicolumn{4}{c}{Radiant Data} \\ 
\hline \hline      
& Observed & Geocentric & Heliocentric \\ 
\hline 
$\alpha\,$ ($^{\circ}$) & 228.2$\pm$1.5 & 204.0$\pm$1.4 & 226.1$\pm$0.4 \\
$\delta\,$ ($^{\circ}$) & 68.0$\pm$0.2 & 67.6$\pm$0.4 & 16.9$\pm$0.4 \\
V\, (km/s) & 15$\pm$1 & 10$\pm$1.5 & 31$\pm$0.6 \\
\hline
\end{tabular}

\scriptsize
\begin{tabular}{l c c c c c}
\hline                         
\multicolumn{6}{c}{Orbital Parameter} \\ 
\hline \hline      
a\, (AU) & e & q & $\omega$\, ($^{\circ}$) & $\omega$\, ($^{\circ}$) & i\, ($^{\circ}$) \\
\hline
1.15$\pm$0.06 & 0.17$\pm$0.04 & 0.953$\pm$0.008 & 126$\pm$9 & 143.43$\pm10^{-4}$ & 17$\pm$3  \\
\hline
\end{tabular}

\scriptsize
\begin{tabular}{p{0.89cm} p{0.78cm} p{0.8cm} p{0.88cm} p{0.88cm} p{0.65cm} p{0.65cm}}

\hline                        
\multicolumn{7}{c}{SPMN160819} \\ 
\hline \hline
Mag & $h_i\, (km)$ & $h_f\, (km)$ & $V_i\, (km/s)$ & $V_f\, (km/s)$ & $M_i\, (kg)$ & $M_f \,(kg)$ \\
\hline
-16.5$\pm$0.5 & 67$\pm$3 & 23$\pm$3 & \,\,\,15.1$\pm$1 & \,\,\,11$\pm$1 & \,\,\,\,\,2100 & \,\,\,190 \\

\hline
\end{tabular}

\caption{Top: SPMN160819 observed, geocentric and heliocentric radiant and velocities. Middle: SPMN160819 calculated orbital parameters. Bottom: SPMN160819 computed atmospheric trajectory, velocity and mass.}
\label{table:SPMN160819}
\end{table}

\begin{figure}
	\includegraphics[width=\columnwidth]{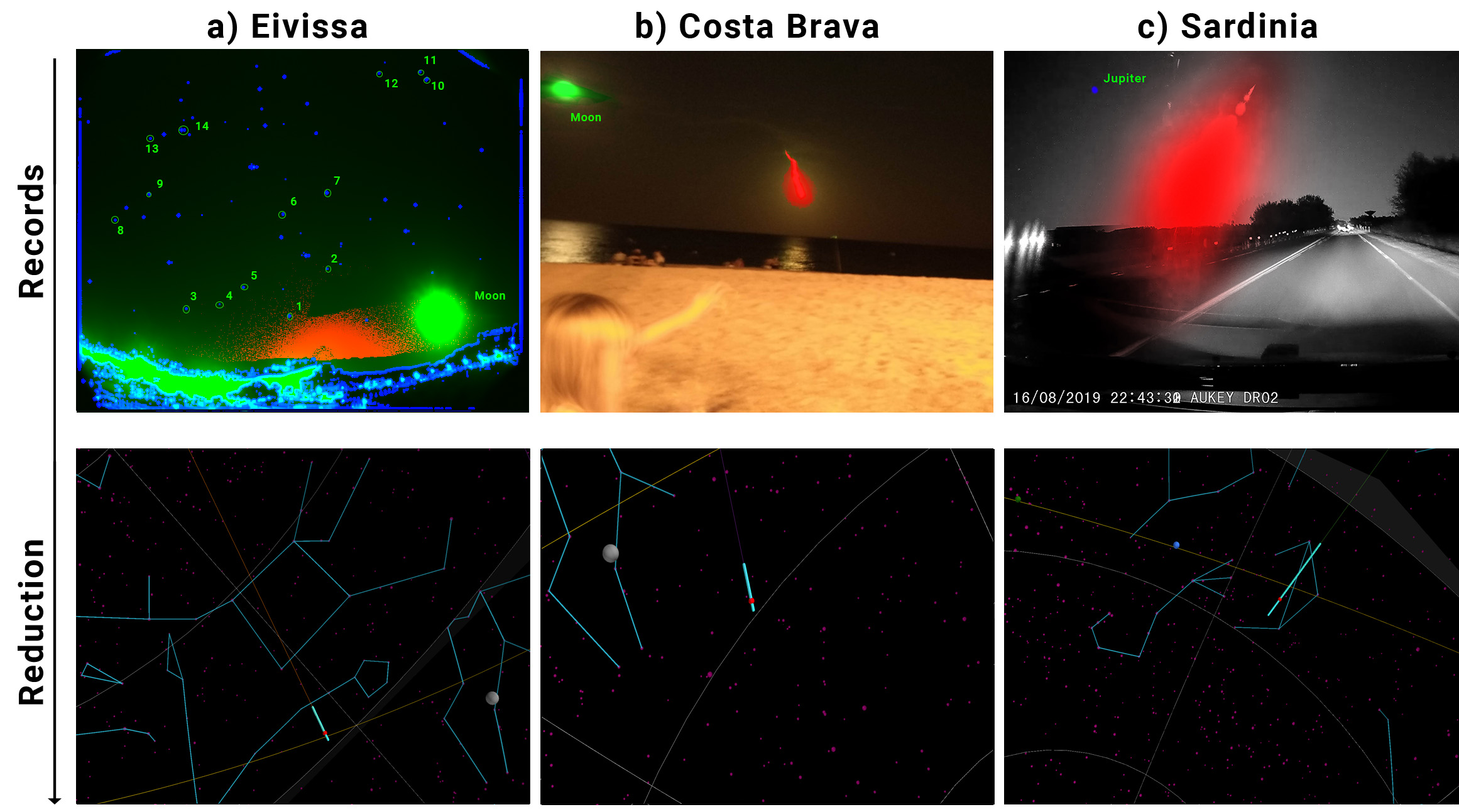}
	\includegraphics[width=\columnwidth]{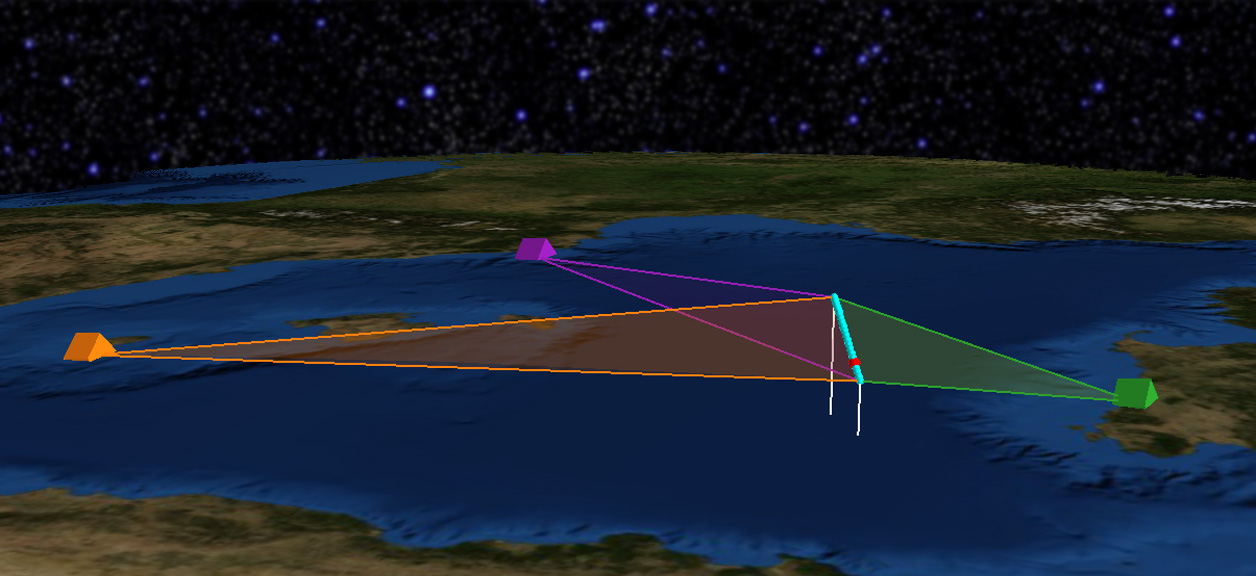}

    \caption{Top: SPMN160819 apparent trajectory recorded and reduced from SPMN Eivissa station (Orange), Sardinia (Green) and Costa Brava (purple). Bottom: SPMN160819 atmospheric trajectory with vertical projection (white). The red dot corresponds to the point of highest radiated energy as registered by CNEOS.}
    \label{fig:figure7}
\end{figure}

Figure~\ref{fig:figure7} shows the stacked frames of the recordings and the graphic representation of the apparent trajectories over the celestial sphere and the atmospheric flight in real scale. Figure~\ref{fig:figure6b} shows the orbit of the SPMN160819 progenitor.

\begin{figure}
	\includegraphics[width=\columnwidth]{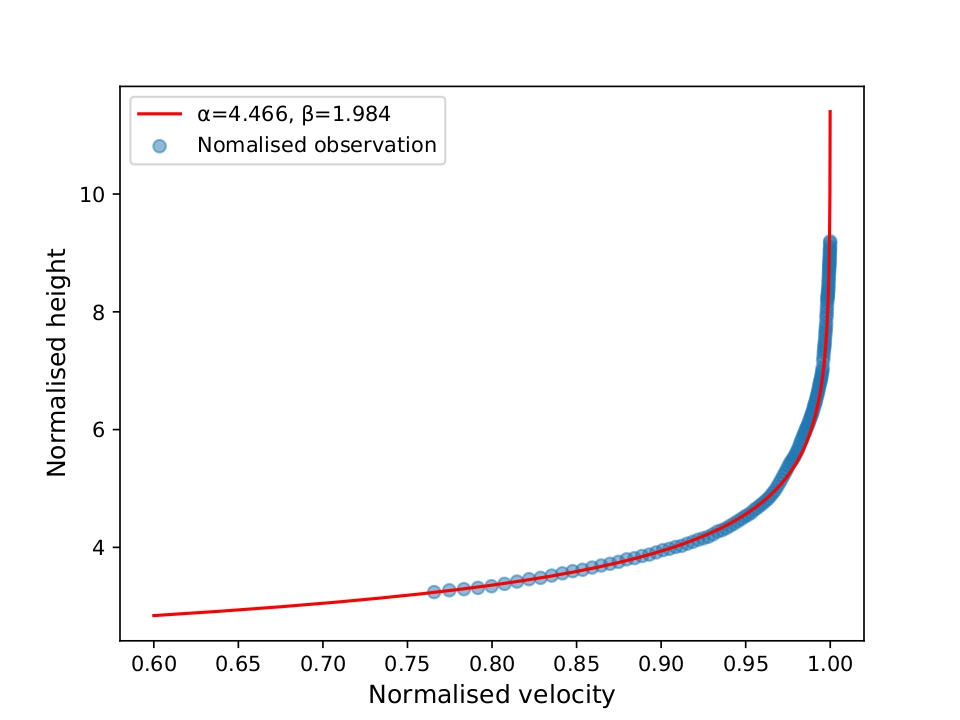}
	\includegraphics[width=\columnwidth]{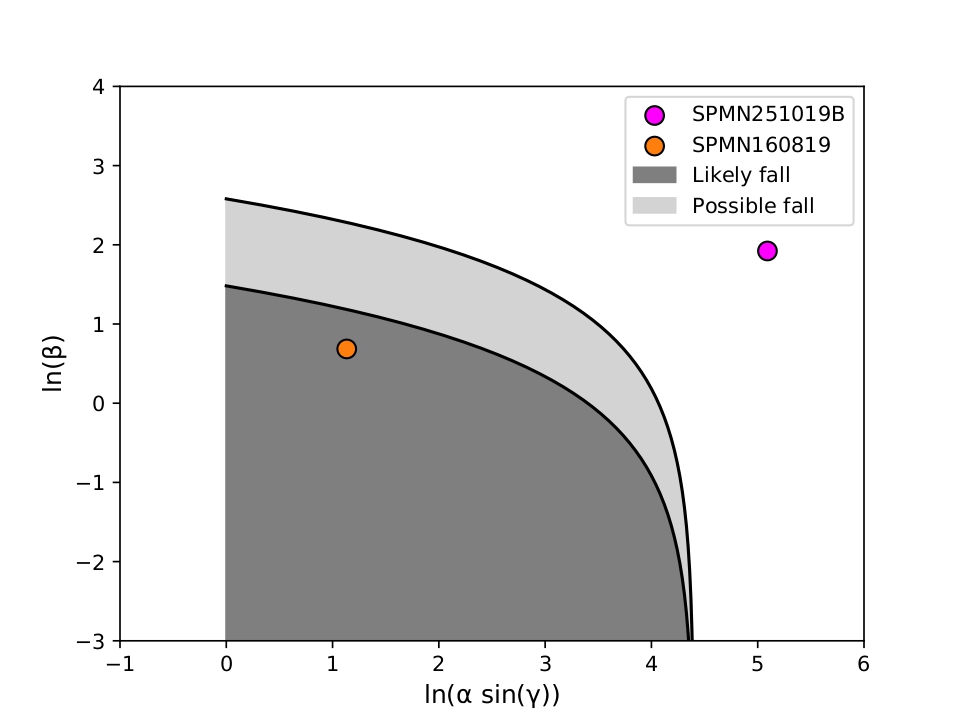}

    \caption{Top: Plot of observational data with velocity normalized to the entry velocity and height normalized to the atmospheric scale height for the SPMN160819 event. Bottom: SPMN160819 flight parametrization. The bounding line for a $50\, g$ meteorite is shown in black for the case where there is no spin ($\mu = 0$) and in grey where spin allows uniform ablation over the entire surface ($\mu = 2/3$). Parameters $\alpha$ and $\beta$ come from Eq. \ref{alphabeta}.}
    \label{fig:figure8}
\end{figure}

The slope between the trajectory and the local horizon is one of the key parameters that define the fate of the meteoroid as a consequence of the ablation. In this case, the trajectory slope was estimated to be $49\,^{\circ}$. After performing the fitting of the normalised velocity and the normalised height in order to parametrize the atmospheric flight (see Figure~\ref{fig:figure8}) and assuming a mean value of ordinary chondrite’s density of $2.7\, g/cm^3$ \citep{CONSOLMAGNOSJ1998, blum2006physics}, a shape change coefficient of $\mu = 2/3$, a shape factor of $A_e=1.3$ and a drag coefficient of $c_d = 1.3$ the masses are calculated from Eq. \ref{Masse}. The initial mass of the meteoroid was estimated to be $2100\, kg$ corresponding to the initial size of $1.2\, m$ and the terminal mass computed is $190\, kg$. Introducing the radiated energy peak recorded by CNEOS, $T_E = 0.089 \,kt $, on Eq. \ref{Diamater} gives a diameter of $1.3 \, m$, which is in good agreement with our results. This emitted energy could be compared with the Villalbeto de la Peña superbolide videotaped on Jan. 4th, 2004 that produced a blast with a kinetic energy of about $0.09 \,kt$ \citep{Llorca2005, trigo2006villalbeto}. The $\alpha - \beta$ criterion shows that this event was likely to produce meteorites, as it is depicted in Figure~\ref{fig:figure8}. The results are shown in Table~\ref{table:SPMN160819}.

\section{Conclusions}

A software tool for the detection and reconstruction of meteor trajectories was developed. The entire reduction procedure is automatic, this way increasing our capacity to quantify, almost in real time, the meteoroid properties, the fireball trajectory, the heliocentric orbit, and its ability to penetrate the atmosphere and pose a potential hazard. Thanks to the application of these new techniques, the analysis of the atmospheric deceleration of cm to m-sized bodies penetrating the atmosphere at hypervelocity is facilitated. In summary, the main conclusions of this work are:
\begin{itemize}
    \item An automatized processing system to perform detection and astrometric reduction of meteor video recordings is presented. The software uses state-of-the-art vision techniques, image processing and motion detection methods to achieve fast astrometry, and a reliable calculation of astrometric errors. 

    \item New reduction techniques allows for avoiding false positives associated with bright flares experienced during the ablation process, as the Kalman filter is implemented to predict the motion of the object in the image. In addition, to discard incorrect points, a post-processing treatment was developed using clustering algorithms.

    \item A corner algorithm is applied to automatically identify reference stars. Subsequent treatment is performed to avoid possible false positives due to presence of other objects, such as trees or buildings. Also, corrections of the atmospheric extinction and refraction, as well as the light aberration due to the Earth’s motion, was implemented to improve the photometry. 

    \item We implemented a model to approximate the distortion of the lenses produced by wide-field and all-sky cameras using quadratic expressions and the simplex algorithm. The software characterizes the meteor ﬂight and computes the pre-atmospheric mass. It identifies if a bolide is a meteorite-dropper using the $\alpha - \beta$ criterion.

    \item A realistic atmospheric trajectory model in 3D was developed. Astrometric errors are propagated to infer the uncertainty in the determination of the heights, radiant and inferred velocity along the luminous path of the fireballs. 

    \item The performance of the software is demostrated by computing the heliocentric orbits of the two study cases. We found that the SPMN251019B fireball data are in good agreement with the values associated with the Southern Taurid meteoroid stream.

    \item In reference to the superbolide SPMN160819, the obtained data show that it was produced by a m-sized sporadic meteoroid that, after disruption in the atmosphere, might have produced meteorites. 

\end{itemize}

\section*{Acknowledgements}

JMT-R, EPA and AR acknowledge financial support from the Spanish Ministry (PGC2018-097374-B-I00 funded by MCI-AEI-FEDER, PI: JMT-R; CTQ2017-89132-P,PI: AR). MG acknowledges support from the Academy of Finland project no. 325806, and the Russian Foundation for Basic Research, project nos. 18-08-00074 and 19-05-00028. AR is indebted to the “Ramón y Cajal” program and DIUE (project 2017SGR1323). This project has received funding from the European Research Council (ERC) under the European Union’s Horizon 2020 research and innovation programme (grant agreement No. 865657) for the project “Quantum Chemistry on Interstellar Grains” (QUANTUMGRAIN). We thank Prof. José A. Docobo and Dr. Pedro P. Campo for the video obtained to exemplify the software (Fig. \ref{fig:figure1}-\ref{fig:figure3}) recorded from Observatorio Astronómico Ramón Maria Aller (OARMA), Universidad de Santiago de Compostela. The casual video recording from Sardinia was kindly provided by Claudio Porcu (Fig. \ref{fig:figure7}). We also thank Quico Terradelles for provinding the persistent train of the SPMN160819 superbolide recorded from Platja d'Aro, Costa Brava (Fig. \ref{fig:figure7}). We also thank preliminary feedback received from that superbolide by Peter Brown (UWO). The authors thank Dr. Eleanor K. Sansom for the detailed and constructive review that helped us to improve this paper.

\section*{Availability of data}
The data underlying this article will be shared on reasonable request to the corresponding author.



\bibliographystyle{mnras}
\bibliography{example} % if your bibtex file is called example.bib

\begin{thebibliography}{}
\makeatletter
\relax
\def\mn@urlcharsother{\let\do\@makeother \do\$\do\&\do\#\do\^\do\_\do\%\do\~}
\def\mn@doi{\begingroup\mn@urlcharsother \@ifnextchar [ {\mn@doi@}
  {\mn@doi@[]}}
\def\mn@doi@[#1]#2{\def\@tempa{#1}\ifx\@tempa\@empty \href
  {http://dx.doi.org/#2} {doi:#2}\else \href {http://dx.doi.org/#2} {#1}\fi
  \endgroup}
\def\mn@eprint#1#2{\mn@eprint@#1:#2::\@nil}
\def\mn@eprint@arXiv#1{\href {http://arxiv.org/abs/#1} {{\tt arXiv:#1}}}
\def\mn@eprint@dblp#1{\href {http://dblp.uni-trier.de/rec/bibtex/#1.xml}
  {dblp:#1}}
\def\mn@eprint@#1:#2:#3:#4\@nil{\def\@tempa {#1}\def\@tempb {#2}\def\@tempc
  {#3}\ifx \@tempc \@empty \let \@tempc \@tempb \let \@tempb \@tempa \fi \ifx
  \@tempb \@empty \def\@tempb {arXiv}\fi \@ifundefined
  {mn@eprint@\@tempb}{\@tempb:\@tempc}{\expandafter \expandafter \csname
  mn@eprint@\@tempb\endcsname \expandafter{\@tempc}}}

\bibitem[\protect\citeauthoryear{{Andreev}}{{Andreev}}{1990}]{Andreev1990}
{Andreev} G.,  1990, in Proceedings of the International Meteor Conference, 9th
  IMC, Violau, Germany, 1990. pp 25--27

\bibitem[\protect\citeauthoryear{Bannister, Boucheron  \& Voelz}{Bannister
  et~al.}{2013}]{bannister2013numerical}
Bannister S.~M.,  Boucheron L.~E.,   Voelz D.~G.,  2013, Publications of the
  Astronomical Society of the Pacific, 125, 1108

\bibitem[\protect\citeauthoryear{Barghini, Gardiol, Carbognani  \&
  Mancuso}{Barghini et~al.}{2019}]{barghini2019astrometric}
Barghini D.,  Gardiol D.,  Carbognani A.,   Mancuso S.,  2019, Astronomy \&
  Astrophysics, 626, A105

\bibitem[\protect\citeauthoryear{{Bellot-Rubio}}{{Bellot-Rubio}}{1992}]{Bellot1992}
{Bellot-Rubio} L.,  1992, Introducción a la Teoría Física de los Meteoros.
Interscience Publishers

\bibitem[\protect\citeauthoryear{Bland}{Bland}{2004}]{Bland2004}
Bland P.~A.,  2004, Astronomy \& Geophysics, 45, 5

\bibitem[\protect\citeauthoryear{Blum, Schr{\"a}pler, Davidsson  \&
  Trigo-Rodriguez}{Blum et~al.}{2006}]{blum2006physics}
Blum J.,  Schr{\"a}pler R.,  Davidsson B.~J.,   Trigo-Rodriguez J.~M.,  2006,
  The Astrophysical Journal, 652, 1768

\bibitem[\protect\citeauthoryear{Borovicka}{Borovicka}{1990}]{borovicka1990comparison}
Borovicka J.,  1990, Bulletin of the Astronomical Institutes of Czechoslovakia,
  41, 391

\bibitem[\protect\citeauthoryear{Borovi{\v{c}}ka}{Borovi{\v{c}}ka}{1992}]{Borovivcka1992astrometry}
Borovi{\v{c}}ka J.,  1992, PAICz, 79

\bibitem[\protect\citeauthoryear{Borovicka, Spurny  \& Keclikova}{Borovicka
  et~al.}{1995}]{Borovicka1995new}
Borovicka J.,  Spurny P.,   Keclikova J.,  1995, Astronomy and Astrophysics
  Supplement Series, 112, 173

\bibitem[\protect\citeauthoryear{Boslough \& Crawford}{Boslough \&
  Crawford}{2008}]{Boslough2008}
Boslough M.,  Crawford D.,  2008, \mn@doi [International Journal of Impact
  Engineering] {10.1016/j.ijimpeng.2008.07.053}, 35, 1441

\bibitem[\protect\citeauthoryear{Bouquet, Baratoux, Vaubaillon, Gritsevich,
  Mimoun, Mousis  \& Bouley}{Bouquet et~al.}{2014}]{Bouquet2014}
Bouquet A.,  Baratoux D.,  Vaubaillon J.,  Gritsevich M.~I.,  Mimoun D.,
  Mousis O.,   Bouley S.,  2014, \mn@doi [Planetary and Space Science]
  {10.1016/j.pss.2014.09.001}, 103, 238

\bibitem[\protect\citeauthoryear{Bradski \& Kaehler}{Bradski \&
  Kaehler}{2000}]{opencv_library}
Bradski G.,  Kaehler A.,  2000, Dr. Dobb's Journal of Software Tools, 25, 120

\bibitem[\protect\citeauthoryear{Brown, Revelle, Tagliaferri  \&
  Hildebrand}{Brown et~al.}{2002a}]{brown2002entry}
Brown P.~G.,  Revelle D.~O.,  Tagliaferri E.,   Hildebrand A.~R.,  2002a,
  Meteoritics \& Planetary Science, 37, 661

\bibitem[\protect\citeauthoryear{Brown, Spalding, ReVelle, Tagliaferri  \&
  Worden}{Brown et~al.}{2002b}]{brown2002flux}
Brown P.,  Spalding R.,  ReVelle D.~O.,  Tagliaferri E.,   Worden S.,  2002b,
  Nature, 420, 294

\bibitem[\protect\citeauthoryear{{Ceplecha}}{{Ceplecha}}{1957}]{Ceplecha1957}
{Ceplecha} Z.,  1957, Bulletin of the Astronomical Institutes of
  Czechoslovakia, \href {https://ui.adsabs.harvard.edu/abs/1957BAICz...8...51C}
  {8, 51}

\bibitem[\protect\citeauthoryear{{Ceplecha}}{{Ceplecha}}{1987}]{Ceplecha1987}
{Ceplecha} Z.,  1987, Bulletin of the Astronomical Institutes of
  Czechoslovakia, \href {https://ui.adsabs.harvard.edu/abs/1987BAICz..38..222C}
  {38, 222}

\bibitem[\protect\citeauthoryear{Colas et~al.,}{Colas et~al.}{2015}]{Colas2014}
Colas F.,  et~al., 2015, in Proceedings International Meteor Conference. pp
  34--38

\bibitem[\protect\citeauthoryear{Colas, Zanda, Bouley, Jeanne, Malgoyre, Birlan
   et~al.}{Colas et~al.}{2020}]{Colas2020}
Colas F.,  Zanda B.,  Bouley S.,  Jeanne S.,  Malgoyre A.,  Birlan M.,
  et~al., 2020, \mn@doi [Submitted to A{\&}A] {10.1051/0004-6361/202038649}

\bibitem[\protect\citeauthoryear{{Consolmagno} \& {Britt}}{{Consolmagno} \&
  {Britt}}{1998}]{CONSOLMAGNOSJ1998}
{Consolmagno} S.,  {Britt} D.~T.,  1998, Meteoritics \& Planetary Science, 33,
  1231

\bibitem[\protect\citeauthoryear{Devillepoix et~al.,}{Devillepoix
  et~al.}{2020}]{Devillepoix2020}
Devillepoix H.,  et~al., 2020, \mn@doi [Planetary and Space Science]
  {10.1016/j.pss.2020.105036}, 191, 105036

\bibitem[\protect\citeauthoryear{Dmitriev, Lupovka  \& Gritsevich}{Dmitriev
  et~al.}{2015}]{Dmitriev2015orbit}
Dmitriev V.,  Lupovka V.,   Gritsevich M.,  2015, Planetary and Space Science,
  117, 223

\bibitem[\protect\citeauthoryear{Drolshagen et~al.,}{Drolshagen
  et~al.}{2020}]{drolshagen2020luminous}
Drolshagen E.,  et~al., 2020, Luminous efficiency based on FRIPON meteors
  (\mn@eprint {arXiv} {2011.06805})

\bibitem[\protect\citeauthoryear{Dubiago}{Dubiago}{1961}]{dubiago1961determination}
Dubiago A.,  1961, New York

\bibitem[\protect\citeauthoryear{Ester, Kriegel, Sander, Xu  et~al.}{Ester
  et~al.}{1996}]{Ester1996}
Ester M.,  Kriegel H.-P.,  Sander J.,  Xu X.,   et~al., 1996, in Kdd. AAAI
  Press, pp 226--231

\bibitem[\protect\citeauthoryear{Gardiol, Cellino  \& Di~Martino}{Gardiol
  et~al.}{2016}]{Gardiol2016}
Gardiol D.,  Cellino A.,   Di~Martino M.,  2016. In Proceedings of the
  International Meteor Conference, Egmond., pp 76--79

\bibitem[\protect\citeauthoryear{{Green}}{{Green}}{1992}]{Green1992}
{Green} D. W.~E.,  1992, International Comet Quarterly, \href
  {https://ui.adsabs.harvard.edu/abs/1992ICQ....14...55G} {14, 55}

\bibitem[\protect\citeauthoryear{Gritsevich}{Gritsevich}{2008a}]{Gritsevich2008pribram}
Gritsevich M.,  2008a, Solar System Research, 42, 372

\bibitem[\protect\citeauthoryear{Gritsevich}{Gritsevich}{2008b}]{Gritsevich2008estimating}
Gritsevich M.,  2008b, in Doklady Physics. p.~588

\bibitem[\protect\citeauthoryear{Gritsevich}{Gritsevich}{2009}]{Gritsevich2009determination}
Gritsevich M.,  2009, Advances in Space Research, 44, 323

\bibitem[\protect\citeauthoryear{Gritsevich \& Koschny}{Gritsevich \&
  Koschny}{2011}]{Gritsevich2011}
Gritsevich M.,  Koschny D.,  2011, \mn@doi [Icarus]
  {10.1016/j.icarus.2011.01.033}, 212, 877

\bibitem[\protect\citeauthoryear{Gritsevich, Stulov  \& Turchak}{Gritsevich
  et~al.}{2009}]{Gritsevich2009classification}
Gritsevich M.~I.,  Stulov V.~P.,   Turchak L.~I.,  2009, \mn@doi [Doklady
  Physics] {10.1134/s1028335809110068}, 54, 499

\bibitem[\protect\citeauthoryear{Gritsevich, Stulov  \& Turchak}{Gritsevich
  et~al.}{2012}]{Gritsevich2012consequences}
Gritsevich M.,  Stulov V.,   Turchak L.,  2012, Cosmic Research, 50, 56

\bibitem[\protect\citeauthoryear{{Gritsevich} et~al.,}{{Gritsevich}
  et~al.}{2014}]{Gritsevich2014}
{Gritsevich} M.,  et~al., 2014, in {Rault} J.~L.,  {Roggemans} P.,  eds,
  Proceedings of the International Meteor Conference, Giron, France, 18-21
  September 2014. pp 162--169

\bibitem[\protect\citeauthoryear{Gritsevich et~al.,}{Gritsevich
  et~al.}{2017}]{Gritsevich2017}
Gritsevich M.,  et~al., 2017, in , Astrophysics and Space Science Proceedings.
Springer International Publishing, pp 153--183,
  \mn@doi{10.1007/978-3-319-46179-3_8}

\bibitem[\protect\citeauthoryear{Gural}{Gural}{2012}]{gural2012new}
Gural P.~S.,  2012, Meteoritics \& Planetary Science, 47, 1405

\bibitem[\protect\citeauthoryear{Harmon \& Nolan}{Harmon \&
  Nolan}{2005}]{HARMON2005}
Harmon J.,  Nolan M.,  2005, \mn@doi [Icarus] {10.1016/j.icarus.2005.01.012},
  176, 175

\bibitem[\protect\citeauthoryear{Hawkes}{Hawkes}{1993}]{Hawkes1993}
Hawkes R.~L.,  1993, in Meteoroids and their Parent Bodies. p.~227

\bibitem[\protect\citeauthoryear{Hoppe}{Hoppe}{1937}]{Hoppe1937}
Hoppe J.,  1937, \mn@doi [Astronomische Nachrichten]
  {10.1002/asna.19372621002}, 262, 169

\bibitem[\protect\citeauthoryear{Jacchia \& Whipple}{Jacchia \&
  Whipple}{1956}]{Jacchia1956}
Jacchia L.~G.,  Whipple F.~L.,  1956, \mn@doi [Vistas in Astronomy]
  {10.1016/0083-6656(56)90021-6}, 2, 982

\bibitem[\protect\citeauthoryear{Jansen-Sturgeon, Sansom, Devillepoix, Bland,
  Towner, Howie  \& Hartig}{Jansen-Sturgeon et~al.}{2020}]{jansen2020dynamic}
Jansen-Sturgeon T.,  Sansom E.~K.,  Devillepoix H.~A.,  Bland P.~A.,  Towner
  M.~C.,  Howie R.~M.,   Hartig B.~A.,  2020, The Astronomical Journal, 160,
  190

\bibitem[\protect\citeauthoryear{Jeanne et~al.,}{Jeanne
  et~al.}{2019}]{jeanne2019calibration}
Jeanne S.,  et~al., 2019, Astronomy \& Astrophysics, 627, A78

\bibitem[\protect\citeauthoryear{Jenniskens}{Jenniskens}{1998}]{Jenniskens1998}
Jenniskens P.,  1998, \mn@doi [Earth, Planets and Space] {10.1186/bf03352149},
  50, 555

\bibitem[\protect\citeauthoryear{Jenniskens \& Vaubaillon}{Jenniskens \&
  Vaubaillon}{2008}]{Jenniskens2008}
Jenniskens P.,  Vaubaillon J.,  2008, \mn@doi [AJ.]
  {10.1088/0004-6256/136/2/725}, 136

\bibitem[\protect\citeauthoryear{Jenniskens \& de Lignie}{Jenniskens \&
  de~Lignie}{1987}]{Jenniskens1987}
Jenniskens P.,  de Lignie M.,  1987, J. DMS, 9

\bibitem[\protect\citeauthoryear{Jenniskens et~al.,}{Jenniskens
  et~al.}{2016}]{Jenniskens2016}
Jenniskens P.,  et~al., 2016, \mn@doi [Icarus] {10.1016/j.icarus.2015.09.013},
  266, 331

\bibitem[\protect\citeauthoryear{{Kresak}}{{Kresak}}{1978}]{Kresak1978}
{Kresak} L.,  1978, Bulletin of the Astronomical Institutes of Czechoslovakia,
  \href {https://ui.adsabs.harvard.edu/abs/1978BAICz..29..129K} {29, 129}

\bibitem[\protect\citeauthoryear{{Levin}}{{Levin}}{1956}]{Levin1956}
{Levin} B.,  1956, Akad. Nauk SSSR, Moscow

\bibitem[\protect\citeauthoryear{Llorca et~al.,}{Llorca
  et~al.}{2005}]{Llorca2005}
Llorca J.,  et~al., 2005, \mn@doi [Meteoritics and Planetary Science]
  {10.1111/j.1945-5100.2005.tb00155.x}, 40, 795

\bibitem[\protect\citeauthoryear{Lyytinen \& Gritsevich}{Lyytinen \&
  Gritsevich}{2016}]{lyytinen2016implications}
Lyytinen E.,  Gritsevich M.,  2016, Planetary and Space Science, 120, 35

\bibitem[\protect\citeauthoryear{Madiedo \& Trigo-Rodr{\'{\i}}guez}{Madiedo \&
  Trigo-Rodr{\'{\i}}guez}{2007}]{Madiedo2007}
Madiedo J.~M.,  Trigo-Rodr{\'{\i}}guez J.~M.,  2007, \mn@doi [Earth, Moon, and
  Planets] {10.1007/s11038-007-9215-x}, 102, 133

\bibitem[\protect\citeauthoryear{McCrosky \& Posen}{McCrosky \&
  Posen}{1968}]{mccrosky1968special}
McCrosky R.~E.,  Posen A.,  1968, SAO Special Report, 273

\bibitem[\protect\citeauthoryear{Meier, Welten, Riebe, Caffee, Gritsevich,
  Maden  \& Busemann}{Meier et~al.}{2017}]{Meier2017}
Meier M. M.~M.,  Welten K.~C.,  Riebe M. E.~I.,  Caffee M.~W.,  Gritsevich M.,
  Maden C.,   Busemann H.,  2017, \mn@doi [Meteoritics {\&} Planetary Science]
  {10.1111/maps.12874}, 52, 1561

\bibitem[\protect\citeauthoryear{Moilanen, Gritsevich  \& Lyytinen}{Moilanen
  et~al.}{2021}]{Moilanen2021}
Moilanen J.,  Gritsevich M.,   Lyytinen E.,  2021, \mn@doi [MNRAS, in revision]
  {https://doi.org/10.1093/mnras/stab586}

\bibitem[\protect\citeauthoryear{Moreno-Ib{\'a}{\~n}ez, Gritsevich  \&
  Trigo-Rodr{\'\i}guez}{Moreno-Ib{\'a}{\~n}ez et~al.}{2015}]{Moreno2015new}
Moreno-Ib{\'a}{\~n}ez M.,  Gritsevich M.,   Trigo-Rodr{\'\i}guez J.~M.,  2015,
  Icarus, 250, 544

\bibitem[\protect\citeauthoryear{Moreno-Ib{\'{a}}{\~{n}}ez,
  Trigo-Rodr{\'{\i}}guez, Gritsevich  \& Silber}{Moreno-Ib{\'{a}}{\~{n}}ez
  et~al.}{2020}]{Moreno2020}
Moreno-Ib{\'{a}}{\~{n}}ez M.,  Trigo-Rodr{\'{\i}}guez J.~M.,  Gritsevich M.,
  Silber E.~A.,  2020, Not. Roy. Astron. Soc., pp 316--324

\bibitem[\protect\citeauthoryear{Motzkin}{Motzkin}{1952}]{motzkin1952new}
Motzkin T.,  1952, Project... scoop

\bibitem[\protect\citeauthoryear{{Murad} \& {Williams}}{{Murad} \&
  {Williams}}{2002}]{Murad2002}
{Murad} E.,  {Williams} I.~P.,  2002, {Meteors in the Earth's Atmosphere}

\bibitem[\protect\citeauthoryear{{Peña-Asensio} et~al.,}{{Peña-Asensio}
  et~al.}{2020a}]{Eloy2020a}
{Peña-Asensio} E.,  et~al., 2020a, in LPI Contribution, 39th Lunar and
  Planetary Science Conference- No., 1391.

\bibitem[\protect\citeauthoryear{{Peña-Asensio}, {Trigo-Rodriguez}, {Mas-Sanz}
   \& {Ribas}}{{Peña-Asensio} et~al.}{2020b}]{Eloy2020b}
{Peña-Asensio} E.,  {Trigo-Rodriguez} {Mas-Sanz} E.,   {Ribas} J.,  2020b.
  EuroPlanet Science Congress 2020. EPSC2020-459

\bibitem[\protect\citeauthoryear{{Rendtel}}{{Rendtel}}{1993}]{Rendtel1993}
{Rendtel} J.,  1993, Handbook for visual meteor observers (Belgium:
  International Meteor Organization)

\bibitem[\protect\citeauthoryear{{Rublee}, {Rabaud}, {Konolige}  \&
  {Bradski}}{{Rublee} et~al.}{2011}]{Rublee2011}
{Rublee} E.,  {Rabaud} V.,  {Konolige} K.,   {Bradski} G.,  2011, in 2011
  International Conference on Computer Vision. pp 2564--2571

\bibitem[\protect\citeauthoryear{Sansom, Bland, Paxman  \& Towner}{Sansom
  et~al.}{2015}]{Sansom2015}
Sansom E.~K.,  Bland P.,  Paxman J.,   Towner M.,  2015, \mn@doi [Meteoritics
  {\&} Planetary Science] {10.1111/maps.12478}, 50, 1423

\bibitem[\protect\citeauthoryear{Sansom et~al.,}{Sansom
  et~al.}{2019}]{Sansom2019}
Sansom E.~K.,  et~al., 2019, \mn@doi [The Astrophysical Journal]
  {10.3847/1538-4357/ab4516}, 885, 115

\bibitem[\protect\citeauthoryear{Sekanina}{Sekanina}{1998}]{Sekanina1998}
Sekanina Z.,  1998, \mn@doi [Planetary and Space Science]
  {10.1016/s0032-0633(96)00154-7}, 46, 191

\bibitem[\protect\citeauthoryear{Silber, Boslough, Hocking, Gritsevich  \&
  Whitaker}{Silber et~al.}{2018}]{Silber2018physics}
Silber E.~A.,  Boslough M.,  Hocking W.~K.,  Gritsevich M.,   Whitaker R.~W.,
  2018, Advances in Space Research, 62, 489

\bibitem[\protect\citeauthoryear{Southworth \& Hawkins}{Southworth \&
  Hawkins}{1963}]{southworth1963statistics}
Southworth R.,  Hawkins G.,  1963, Smithsonian Contributions to Astrophysics,
  7, 261

\bibitem[\protect\citeauthoryear{{Steyaert}}{{Steyaert}}{1990}]{Steyaert1990}
{Steyaert} C.,  1990, Photographic Astrometry.
International Meteor Organization

\bibitem[\protect\citeauthoryear{{Stulov} et~al.}{{Stulov}
  et~al.}{1995}]{Stulov1995}
{Stulov} V.~P.,  et~al., 1995, Moscow: Nauka, p.~236

\bibitem[\protect\citeauthoryear{Suzuki \& Abe}{Suzuki \&
  Abe}{1985}]{Suzuki1985}
Suzuki S.,  Abe K.,  1985, \mn@doi [Computer Vision, Graphics, and Image
  Processing] {10.1016/0734-189x(85)90016-7}, 30, 32

\bibitem[\protect\citeauthoryear{{Tanbakouei}, {Trigo-Rodr{\'\i}guez}, {Sort},
  {Michel}, {Blum}, {Nakamura}  \& {Williams}}{{Tanbakouei}
  et~al.}{2019}]{Safoura2019}
{Tanbakouei} S.,  {Trigo-Rodr{\'\i}guez} J.~M.,  {Sort} J.,  {Michel} P.,
  {Blum} J.,  {Nakamura} T.,   {Williams} I.,  2019, \mn@doi [\aap]
  {10.1051/0004-6361/201935380}, \href
  {https://ui.adsabs.harvard.edu/abs/2019A&A...629A.119T} {629, A119}

\bibitem[\protect\citeauthoryear{{Tatum}}{{Tatum}}{2019}]{Tatum2019}
{Tatum} J.~B.,  2019, Celestial Mechanics. Chapter 11: Photographic Astrometry.

\bibitem[\protect\citeauthoryear{{Trigo-Rodriguez} \&
  {Williams}}{{Trigo-Rodriguez} \& {Williams}}{2017}]{Trigo2017}
{Trigo-Rodriguez} J.~M.,  {Williams} I.~P.,  2017, Springer, pp 11--32

\bibitem[\protect\citeauthoryear{Trigo-Rodriguez, Llorca, Borovi{\v{c}}ka  \&
  Fabregat}{Trigo-Rodriguez et~al.}{2003}]{Trigo2003}
Trigo-Rodriguez J.~M.,  Llorca J.,  Borovi{\v{c}}ka J.,   Fabregat J.,  2003,
  Meteoritics \& Planetary Science, 38, 1283

\bibitem[\protect\citeauthoryear{Trigo-Rodr{\'\i}guez, Llorca  \&
  Fabregat}{Trigo-Rodr{\'\i}guez et~al.}{2004}]{Trigo2004}
Trigo-Rodr{\'\i}guez J.~M.,  Llorca J.,   Fabregat J.,  2004, Monthly Notices
  of the Royal Astronomical Society, 348, 802

\bibitem[\protect\citeauthoryear{Trigo-Rodriguez et~al.,}{Trigo-Rodriguez
  et~al.}{2005}]{trigo2005a}
Trigo-Rodriguez J.,  et~al., 2005, Earth, Moon, and Planets, 95, 553

\bibitem[\protect\citeauthoryear{Trigo-Rodr{\'i}guez, Borovi{\v{c}}ka,
  Spurn{\`y}, Ortiz, Docobo, Castro-Tirado  \& Llorca}{Trigo-Rodr{\'i}guez
  et~al.}{2006a}]{trigo2006villalbeto}
Trigo-Rodr{\'i}guez J.~M.,  Borovi{\v{c}}ka J.,  Spurn{\`y} P.,  Ortiz J.~L.,
  Docobo J.~A.,  Castro-Tirado A.~J.,   Llorca J.,  2006a, Meteoritics \&
  Planetary Science, 41, 505

\bibitem[\protect\citeauthoryear{Trigo-Rodr{\'{\i}}guez, Llorca, Castro-Tirado,
  Ortiz, Docobo  \& Fabregat}{Trigo-Rodr{\'{\i}}guez et~al.}{2006b}]{Trigo2006}
Trigo-Rodr{\'{\i}}guez J.~M.,  Llorca J.,  Castro-Tirado A.~J.,  Ortiz J.~L.,
  Docobo J.~A.,   Fabregat J.,  2006b, \mn@doi [Astronomy {\&} Geophysics]
  {10.1111/j.1468-4004.2006.47626.x}, 47, 6.26

\bibitem[\protect\citeauthoryear{{Trigo-Rodriguez}, {Madiedo}, {Llorca},
  {Gural}, {Pujols}  \& {Tezel}}{{Trigo-Rodriguez} et~al.}{2007}]{Trigo2007}
{Trigo-Rodriguez} J.~M.,  {Madiedo} J.~M.,  {Llorca} J.,  {Gural} P.~S.,
  {Pujols} P.,   {Tezel} T.,  2007, \mn@doi [\mnras]
  {10.1111/j.1365-2966.2007.11966.x}, \href
  {https://ui.adsabs.harvard.edu/abs/2007MNRAS.380..126T} {380, 126}

\bibitem[\protect\citeauthoryear{Trigo-Rodr{\'{\i}}guez, Madiedo, Williams,
  Castro-Tirado, Llorca, V{\'{\i}}tek  \&
  Jel{\'{\i}}nek}{Trigo-Rodr{\'{\i}}guez et~al.}{2009}]{TrigoRodrguez2009}
Trigo-Rodr{\'{\i}}guez J.~M.,  Madiedo J.~M.,  Williams I.~P.,  Castro-Tirado
  A.~J.,  Llorca J.,  V{\'{\i}}tek S.,   Jel{\'{\i}}nek M.,  2009, \mn@doi
  [Monthly Notices of the Royal Astronomical Society]
  {10.1111/j.1365-2966.2008.14363.x}, 394, 569

\bibitem[\protect\citeauthoryear{Trigo-Rodr{\'\i}guez
  et~al.,}{Trigo-Rodr{\'\i}guez et~al.}{2015}]{Trigo2015orbit}
Trigo-Rodr{\'\i}guez J.~M.,  et~al., 2015, Monthly Notices of the Royal
  Astronomical Society, 449, 2119

\bibitem[\protect\citeauthoryear{Trigo-Rodríguez}{Trigo-Rodríguez}{2019}]{Trigo2019}
Trigo-Rodríguez J.~M.,  2019, in 2053-2563, Hypersonic Meteoroid Entry
  Physics.
IOP Publishing, pp 4--1 to 4--23

\bibitem[\protect\citeauthoryear{Turchak \& Gritsevich}{Turchak \&
  Gritsevich}{2014}]{Turchak2014meteoroids}
Turchak L.~I.,  Gritsevich M.~I.,  2014, Journal of Theoretical and Applied
  Mechanics, 44, 15

\bibitem[\protect\citeauthoryear{Vinkovic \& Gritsevich}{Vinkovic \&
  Gritsevich}{2020}]{Vinkovic2020}
Vinkovic D.,  Gritsevich M.,  2020, \mn@doi [Journal of the Geographical
  Institute Jovan Cvijic, {SASA}] {10.2298/ijgi2001045v}, 70, 45

\bibitem[\protect\citeauthoryear{Welch, Bishop  et~al.}{Welch
  et~al.}{1997}]{welch1997introduction}
Welch G.,  Bishop G.,   et~al., 1997, Chapel Hill, NC, USA

\bibitem[\protect\citeauthoryear{Wells}{Wells}{1986}]{wells1986efficient}
Wells W.~M.,  1986, IEEE Transactions on Pattern Analysis and Machine
  Intelligence, pp 234--239

\bibitem[\protect\citeauthoryear{Weryk, Brown, Domokos, Edwards, Krzeminski,
  Nudds  \& Welch}{Weryk et~al.}{2007}]{Weryk2007}
Weryk R.~J.,  Brown P.~G.,  Domokos A.,  Edwards W.~N.,  Krzeminski Z.,  Nudds
  S.~H.,   Welch D.~L.,  2007, \mn@doi [Earth, Moon, and Planets]
  {10.1007/s11038-007-9183-1}, 102, 241

\bibitem[\protect\citeauthoryear{{Whipple} \& {Jacchia}}{{Whipple} \&
  {Jacchia}}{1957}]{Whipple1957}
{Whipple} F.~L.,  {Jacchia} L.~G.,  1957, Smithsonian Contributions to
  Astrophysics, \href {https://ui.adsabs.harvard.edu/abs/1957SCoA....1..183W}
  {1, 183}

\makeatother
\end{thebibliography}

%\begin{thebibliography}{99}
%\end{thebibliography}

%%%%%%%%%%%%%%%%%%%%%%%%%%%%%%%%%%%%%%%%%%%%%%%%%%

%%%%%%%%%%%%%%%%% APPENDICES %%%%%%%%%%%%%%%%%%%%%

\appendix

\section{Calculations}

\begin{table*}

\caption{SPMN251019B data reduction for the station Eivissa, Folgueroles and Montseny. SAO number, plate coordinates in pixels ($x$, $y$), standard coordinates ($\xi$, $\eta$), right ascension and declination and their respective errors are shown.}
\label{table:SPMN251019Bdatareduction}

\begin{tabular}{l c c c c c c c c c c}
\hline                         
Station & Ref & SAO & x (px) & y (px) & $\xi$ & $\eta$ & RA ($^{\circ}$) & DE ($^{\circ}$) & err. RA ($\%$) & err. DE ($\%$) \\ 
\hline \hline                

Eivissa&1a&30631&457.2&497.85&0.09&0.54&268.302&57.01&0.009&0.009\\
&2a&17074&538.1&390.8&0.31&0.73&245.92&61.57&0.018&0.053\\
&3a&17365&493.6&390.2&0.18&0.78&257.125&65.769&0.067&0.06\\
&4a&18222&409.2&404.0&-0.09&0.81&287.974&67.721&0.023&0.04\\
&5a&8220&511.9&305.0&.35&1.13&230.092&71.859&0.032&0.032\\
&6a&8102&504.7&282.2&0.37&1.27&222.576&74.174&0.017&0.023\\
&7a&19019&333.85&432.5&-0.34&0.76&310.523&61.908&0.03&0.094\\
&8a&34137&257.6&411.3&-0.65&0.9&332.736&58.253&0.049&0.01\\
&9a&19302&311.65&414.0&-0.43&0.85&319.53&62.642&0.116&0.061\\
&10a&20268&276.5&345.85&-0.62&1.23&342.449&66.23&0.009&0.012\\
\hline
Folgueroles&1b&60198&393.05&45.75&0.57&-0.1&113.477&31.89&0.034&0.076\\
&2b&79666&405.2&91.35&0.56&-0.18&116.042&28.03&0.037&0.078\\
&3b&95895&614.1&48.65&1.16&-0.35&99.092&15.758&0.004&<0.001\\
&4b&115456&575.5&210.75&0.9&-0.64&111.678&8.299&0.029&0.024\\
&5b&115756&574.55&258.85&0.85&-0.75&114.641&5.236&0.024&0.026\\
&6b&61414&210.5&185.1&0.13&-0.14&140.163&34.395&0.012&0.063\\
&7b&81064&201.95&302.25&0.02&-0.31&148.141&26.014&0.038&<0.001\\
&8b&81004&226.85&313.9&0.04&-0.35&146.409&23.782&0.066&0.103\\
&9b&98967&241.15&442.5&-0.06&-0.6&152.053&11.981&0.012&0.037\\
\hline
Montseny&1c&94027&386.1&234.85&0.15&-0.24&68.982&16.52&0.014&0.062\\
&2c&77168&363.4&420.8&-0.06&-0.01&81.566&28.612&0.078&0.453\\
&3c&39955&261.9&466.0&0.02&0.23&75.488&43.823&0.086&0.285\\
&4c&40186&262.5&496.7&-0.02&0.28&79.165&45.996&0.02&0.136\\
&5c&58636&363.8&510.75&-0.18&0.12&89.914&37.213&0.06&0.455\\
&6c&40750&308.25&547.6&-0.16&0.28&89.864&44.945&0.021&0.101\\
&7c&40756&299.9&553.75&-0.15&0.3&89.965&45.934&0.042&0.306\\
&8c&39053&154.8&400.15&0.28&0.36&55.739&47.789&0.028&0.037\\
&9c&38787&124.0&397.95&0.33&0.42&51.091&49.863&0.038&0.09\\
&10c&23789&87.95&410.15&0.37&0.52&46.212&53.509&0.033&0.052\\

\hline
\end{tabular}
\end{table*}

\begin{table*}
\caption{SPMN160819 data reduction for the station Eivissa. SAO number, plate coordinates in pixels ($x$, $y$), standard coordinates ($\xi$, $\eta$), right ascension and declination and their respective errors are shown.}
\label{table:SPMN160819datareduction}
\begin{tabular}{l c c c c c c c c c}
\hline                         
Ref & SAO & x (px) & y (px) & $\xi$ & $\eta$ & RA ($^{\circ}$) & DE ($^{\circ}$) & err. RA ($\%$) & err. DE ($\%$) \\ 
\hline \hline                

1&91781&339.0&424.8&-0.54&-0.32&3.309&15.184&0.416&0.541\\
2&108378&400.05&349.75&-0.26&-0.39&346.19&15.205&0.395&0.327\\
3&54471&175.55&413.15&-0.68&0.15&17.433&35.621&0.186&0.091\\
4&54058&228.2&406.25&-0.61&0.01&9.832&30.861&0.194&0.122\\
5&73765&266.35&379.4&-0.5&-0.06&2.097&29.09&0.464&0.101\\
6&90981&326.7&299.65&-0.21&-0.13&345.944&28.083&0.279&0.142\\
7&90816&357.5&293.5&-0.16&-0.2&342.501&24.602&0.18&0.04\\
8&90734&329.2&262.85&-0.11&-0.09&340.751&30.221&0.026&0.363\\
9&90238&399.05&229.1&0.06&-0.23&331.753&25.345&0.618&1.342\\
10&127029&492.05&240.0&0.15&-0.49&326.046&9.875&0.179&0.06\\
11&22268&62.45&272.6&-0.46&0.62&21.454&60.235&0.652&0.283\\
12&11482&79.75&252.75&-0.39&0.59&14.177&60.717&0.182&0.08\\
13&21609&107.7&267.8&-0.39&0.5&10.127&56.537&0.153&0.258\\
14&21133&116.65&230.6&-0.29&0.51&2.295&59.15&0.281&0.365\\
15&125122&557.9&48.2&0.75&-0.45&297.696&8.868&0.133&0.474\\
16&105223&547.6&36.1&0.77&-0.42&296.565&10.613&0.138&0.227\\
17&105500&482.4&38.6&0.67&-0.26&299.689&19.492&0.347&0.776\\
18&20268&118.45&142.55&-0.08&0.58&342.42&66.2&0.342&0.072\\
19&34137&168.25&129.25&0.01&0.46&332.714&58.201&0.162&1.578\\
20&19302&166.85&81.7&0.12&0.5&319.645&62.586&0.186&0.166\\
21&19019&178.55&53.35&0.2&0.49&311.322&61.839&0.291&0.582\\

\hline
\end{tabular}
\end{table*}

%%%%%%%%%%%%%%%%%%%%%%%%%%%%%%%%%%%%%%%%%%%%%%%%%%

% Don't change these lines
\bsp	% typesetting comment
\label{lastpage}
\end{document}